\documentclass[prb,twocolumn,showpacs,amsmath,amssymb,superscriptaddress]{revtex4-1}
\usepackage{graphicx}
\usepackage{color}
\usepackage{braket}
\usepackage{amsmath}
\usepackage{amsfonts}
\usepackage{MnSymbol}
\usepackage{amsmath,amssymb,mathrsfs,ulem}
\usepackage[colorlinks = true,linkcolor = blue,urlcolor  = blue,     citecolor = blue,anchorcolor = blue]{hyperref}
\usepackage{bbm}
\usepackage[dvipsnames]{xcolor}

\definecolor{amber}{rgb}{1.0, 0.75, 0.0}

\date{\today}                  

\begin{document}

\title{Universal properties of boundary and interface charges in multichannel one-dimensional models without symmetry constraints}

\author{Niclas M\"uller}
\affiliation{Institut f\"ur Theorie der Statistischen Physik, RWTH Aachen, 
52056 Aachen, Germany and JARA - Fundamentals of Future Information Technology}

\author{Kiryl Piasotski}
\affiliation{Institut f\"ur Theorie der Statistischen Physik, RWTH Aachen, 
52056 Aachen, Germany and JARA - Fundamentals of Future Information Technology}

\author{Dante M. Kennes}
\affiliation{Institut f\"ur Theorie der Statistischen Physik, RWTH Aachen, 
52056 Aachen, Germany and JARA - Fundamentals of Future Information Technology}
\affiliation{Max Planck Institute for the Structure and Dynamics of Matter, Center for Free Electron Laser Science, 22761 Hamburg, Germany}

\author{Herbert Schoeller}
\affiliation{Institut f\"ur Theorie der Statistischen Physik, RWTH Aachen, 
52056 Aachen, Germany and JARA - Fundamentals of Future Information Technology}

\author{Mikhail Pletyukhov}
\email[Email: ]{pletmikh@physik.rwth-aachen.de}
\affiliation{Institut f\"ur Theorie der Statistischen Physik, RWTH Aachen, 
52056 Aachen, Germany and JARA - Fundamentals of Future Information Technology}

\begin{abstract}
The boundary charge that accumulates at the edge of a one-dimensional single-channel insulator is known to possess the universal property, that its change under a lattice shift towards the edge by one site is given by the sum of the average bulk electronic density and a topologically invariant contribution, restricted to the values $0$ and $-1$ [Phys. Rev. B \textbf{101}, 165304 (2020)]. This quantized contribution is associated with particle-hole duality, ensures charge conservation and fixes the mod(1) ambiguity appearing in the Modern Theory of Polarization. In the present work we generalize the above-mentioned single-channel results to the multichannel case by employing the technique of boundary Green's functions. We show that the topological invariant associated with the change in boundary charge under a lattice shift in multichannel models can be expressed as a winding number of a certain combination of components of bulk Green's functions as function of the complex frequency, as it encircles the section of the energy axis that corresponds to the occupied part of the spectrum. We observe that this winding number is restricted to values ranging from $-N_c$ to $0$, where $N_c$ is the number of channels (orbitals) per site. Furthermore, we consider translationally invariant one-dimensional multichannel models with an impurity and introduce topological indices which correspond to the quantized charge that accumulates around said impurity. These invariants are again given in terms of winding numbers of combinations of components of bulk Green's functions. Through this construction we provide a rigorous mathematical proof of the so called \textit{nearsightedness principle} formulated by W.~Kohn [Phys. Rev. Lett. \textbf{76}, 3168 (1996)] for noninteracting multichannel lattice models. 

\end{abstract}


\maketitle

\section{Introduction}
Over the last decades, the study of topological insulators, sparked by the discoveries of the quantum Hall effect [\onlinecite{Klitzing_1980}, \onlinecite{Thouless_1982}] and later of the quantum spin Hall effect [\onlinecite{Konig_2007}], has attracted much interest. Topological insulators (TI) are states of matter with a gapped bulk and symmetry-protected gapless edge states [\onlinecite{He_2019}]. These dissipationless edge states are at the center of interest in the field of TIs, due to their numerous promising applications in quantum computing [\onlinecite{Bomantara_2018, Mong_2014, Stern_2013, Miyake_2010}] and spintronics [\onlinecite{He_2019}, \onlinecite{Pesin_2012}], and their exponential localization poses questions concerning the nature of the charge distribution in their spatial vicinity [\onlinecite{pletyukhov_etal_prb_20}]. As particularly interesting in this respect appear both the boundary charge $Q_{B}$, i.e.~the charge that accumulates at the edge of a finite insulator, and the interface charge $Q_I$, i.e.~the charge that accumulates around the interface shared by a pair of insulators [\onlinecite{Pletyukhov_etal3_2020}]. The study of both of these quantities has a long history [\onlinecite{jackiw_rebbi_prd_76,su_schrieffer_heeger,jackiw_schrieffer_nuclphys_81,rice_mele_prl_82,kivelson_prb_83,su_schrieffer_prl_81,heeger_etal_review_RMP_88,witten_physlett_79,takayama_etal_prb_80,goldstone_wilczek_prl_81,jackiw_semenoff_prl_83}],
connecting them to the field of TIs however is a rather young endeavor. 
\par
With the emergence of the so-called modern theory of polarization (MTP), interest in the study of the boundary charge reawakened [\onlinecite{King_1993, Vanderbilt_1993, Resta_1994, Kudin_2007, Marzari_2012, Spaldin-2012, Rhim_2017, Miert}]. One of the major achievements of the MTP lies in the so-called surface charge theorem, relating the surface (boundary) charge to the bulk polarization, which in turn is related  [\onlinecite{King_1993}, \onlinecite{Vanderbilt_1993}] to the Zak-Berry phase [\onlinecite{zak_89}]. However, the MTP is restricted by the fact that the Zak-Berry phase is defined modulo an integer, since upon an Abelian gauge transformation it changes by the winding number of the corresponding phase. This fact complicates bridging the field of TIs with MTP, since MTP is unable to predict the number of edge states, the central quantity in the field of TIs.
\par
Topological indices related to the boundary charge in one-dimensional, single channel models without symmetry constraints were introduced recently [\onlinecite{pletyukhov_etal_prb_20}, \onlinecite{Pletyukhov_etal2_2020}]. In particular it was shown, that knowledge of the exact eigenstates of a given model allows to find the gauge in which there is a unique relation between the boundary charge $Q_{B,\alpha}$ of a given band $\alpha$ and the corresponding Zak-Berry phase, fixing the unknown integer of the surface charge theorem. A new topological invariant underpinning the universal behavior of the boundary charge upon continuous shifting of the lattice towards the boundary was introduced [\onlinecite{pletyukhov_etal_prb_20}, \onlinecite{Pletyukhov_etal2_2020}]. It was shown, that the boundary charge $Q_{B}$ is connected to universal long-wavelength properties of topological insulators, making the study of the boundary charge an invaluable tool for the characterization of TIs beyond symmetry constraints [\onlinecite{pletyukhov_etal_prb_20, Park_etal_2016, Thakurathi_etal_2018, Pletyukhov_etal2_2020, Lin_etal_2020, Weber_etal_2021, Lin_etal_2021, Laubscher_etal_2021}]. 
\par
In another recent work [\onlinecite{Pletyukhov_etal3_2020}] the universal properties of boundary and interface charges were embedded in a general framework, relating them generically to the nearsightedness principle [\onlinecite{Kohn_1996}, \onlinecite{Prodan_Kohn_2005}]. This principle states a fundamental property of insulators that local perturbations by external fields lead only to local charge redistributions resulting in an addition/removal of an integer number of electrons to/from a perturbed region. Based on this theorem two invariants were established, with quantized values in generic systems, including interactions, random disorder, and multi-channel systems. The first invariant is related to the change of the boundary charge by discrete translations of the lattice, similar to the one described above for single-channel systems. The second invariant is related to the sum of the two boundary charges left and right to a barrier separating a translational invariant lattice in two subparts. The quantization of this invariant is related to a local inversion of the lattice and, together with the first invariant, provides the basis for the quantization of interface charges at domain walls  [\onlinecite{jackiw_rebbi_prd_76,su_schrieffer_heeger,jackiw_schrieffer_nuclphys_81,rice_mele_prl_82,kivelson_prb_83,su_schrieffer_prl_81,heeger_etal_review_RMP_88}] and the generalization of the Goldstone-Wilczek formula [\onlinecite{witten_physlett_79,takayama_etal_prb_80,goldstone_wilczek_prl_81,jackiw_semenoff_prl_83}] to arbitrary tight-binding lattices.
\par
The topic of the present paper is to analyse the two invariants introduced in Ref.~[\onlinecite{Pletyukhov_etal3_2020}] in all detail for noninteracting multi-channel tight-binding models. Via an explicit representation in terms of a winding number defined purely in terms of bulk quantities of the infinite system, we rigorously prove the quantization and provide a mean to explicitly calculate their integer values from bulk properties. This generalizes the results of Refs.~[\onlinecite{pletyukhov_etal_prb_20}, \onlinecite{Pletyukhov_etal2_2020}] to systems with more than a single orbital per lattice site, establishes a bulk-boundary correspondence between universal properties of the boundary charge and bulk properties, and provides a rigorous proof of the nearsightedness principle.
\par
To calculate the invariants we use the method of boundary Green's functions (BGF) [\onlinecite{Arrachea_etal_2009, Zazunov_etal_2016, Peng_etal_2017, Zazunov_etal_2017, Komnik_etal_2017, Zazunov_etal_2018, Pinon_etal_2020, Alvarado_etal_2020}]. Complementary to the method of [\onlinecite{pletyukhov_etal_prb_20}] this technique allows for the direct construction of the open boundary lattice Green's function, avoiding the computation of eigenstates [\onlinecite{Komnik_etal_2017}]. The BGF method proved to be useful in analyzing electronic transport in superconducting systems [\onlinecite{Zazunov_etal_2017}, \onlinecite{Zazunov_etal_2018}], thermal transport in spin heterostructures [\onlinecite{Arrachea_etal_2009}], as well as transport in topological superconductors hosting Majorana bound states [\onlinecite{Zazunov_etal_2016}, \onlinecite{Alvarado_etal_2020}]. Lately, the boundary Green's function technique was generalized to higher dimensional systems [\onlinecite{Pinon_etal_2020}] where the boundaries manifest as lines and planes in two and three dimensions, respectively. Other noteworthy applications of the BGF formalism include the efficient construction of topological phase diagrams [\onlinecite{Peng_etal_2017}], as well as bulk-boundary correspondence related investigations in both non-interacting [\onlinecite{Essin_Gurarie_2011}] and interacting topological insulators [\onlinecite{Gurarie_2011}]. Similar methods were also extensively used in the 60s and 70s with respect to  determining the electronic spectrum as well as conductivity in one-dimensional metallic systems with strong disorder [\onlinecite{bychkov_1966,Bychkov_1967,Kazamanyan_19721}].
\par
In the present work the method of boundary Green's functions is used to study boundary  and interface charges and their associated topological invariants in multichannel extensions of generalized Aubry-Andr\'e-Harper models [\onlinecite{aa_model}, \onlinecite{harper_model}]. Employing the BGF technique, we express the boundary charge of a one-dimensional multichannel insulator in terms of an energy integral of the local spectral density which, in turn, is expressed via the bulk Green's functions. Such a representation is beneficial since it avoids the direct diagonalization of the semi-infinite system's Hamiltonian. We introduce the topological invariant $I$ associated with the change of boundary charge $\Delta{Q}_{B}$ (relative to the average charge per site in the bulk $\bar{\rho}$) under lattice translations. This invariant is expressed as a winding number of a particular combination of components of bulk Green's functions, clearly demonstrating the bulk-boundary correspondence discussed within the realm of topological insulators [\onlinecite{Fidkowski_etal_2011, Mong_Shivamoggi_2011, Fukui_etal_2012, Yu_etal_2017}]. When the whole lattice is shifted by a single site some number of edge states (each contributing a unit of electron charge) may either cross the chemical potential from above or below respectively, hence maintaining the integer valuedness of this invariant. Similar arguments are frequently used in discussions of adiabatic charge pumping [\onlinecite{Thouless_83}, \onlinecite{Niu_Thouless_1984}]. As opposed to the single channel case, where this invariant is limited to the values $0$ and $-1$, in a system with $N_{c}$ orbitals per site $I$ can take on integers in the range $-N_c\leq I \leq 0$, naturally generalizing the $N_{c}=1$ case (see Refs. [\onlinecite{pletyukhov_etal_prb_20}, \onlinecite{Pletyukhov_etal2_2020}]). It is worth noting that the form of bulk-boundary correspondence encompassed by this invariant is not limited to any particular symmetry classes and is purely based on such physical principles as charge conservation and Pauli exclusion principle. To the best of our knowledge this invariant was not established rigorously
in the literature and thus provides a novel contribution to the fields of topological insulators, boundary Green's functions and the modern theory of polarization.
\par
We also adopt the BGF formalism to study the quantization of interface charge and establish rigorously another invariant related to the sum of the boundary charges left and right to a barrier. In particular, we study two types of interface models: The first is obtained from a translationally invariant model by adding a finite potential barrier on a given site, whereas the second type is obtained from a translationally invariant model by weakening the link between two adjacent sites. We derive expressions for the interface charge in both impurity and weak link models in terms of winding numbers of combinations of components of bulk Green's functions, thus demonstrating the quantization of the interface charge. Since the unit cells are perfectly matched, there is no dipole moment in the interface charge distribution, which is thus generated solely by removing a number of valence band electrons to the emerging interface-localized bound states. This result is of fundamental importance in the theory of topological insulators, since it provides the direct analytical proof of the nearsightedness principle [\onlinecite{Kohn_1996}, \onlinecite{Prodan_Kohn_2005}]. 
This principle assigns the robustness property to excess charges localized near a boundary or at an interface of two insulators (cf. [\onlinecite{Pletyukhov_etal3_2020}] for the analytic proof of the interface charge quantization in a special case).
\par
Finally, we substantiate our findings with a number of numerical examples, comparing the winding numbers with their physical counterparts (i.e.~boundary and interface charges) as computed from exact diagonalization of finite systems. We use randomly generated Hamiltonians for this comparison, demonstrating the validity of these novel invariants. 
\par
This paper is organized as follows. In Sec. \ref{sec:BGF} we introduce the class of models under consideration and define their boundary Green's functions. In Sec. \ref{sec:boundaryCharge} we define the boundary charge for multichannel one-dimensional models and express it in terms of boundary Green's functions. We then use results from Sec. \ref{sec:BGF} to express the change in boundary charge upon lattice shifts via the winding number of a specific combination of components of bulk Green's functions. In Sec. \ref{sec:interfaceCharge} we define the interface charge for two different impurity models: 1) finite potential on a single site and 2) a weakened link between two adjacent unit cells. We express the interface charge in both cases in terms of boundary Green's functions and use this representation to cast the expression for the interface charge in the form of a winding number. In Sec. \ref{sec:numerics} we explain how to efficiently evaluate the various winding numbers and then show examples demonstrating the validity of all three invariants as applied to randomly generated models. Finally, in Sec. \ref{sec:summary} we state our summary.

\section{Boundary Green's function}
\label{sec:BGF}

Let us consider a class of translationally invariant one-dimensional (1D) lattice models with $Z$ sites per unit cell (labelled by $1 \leq j \leq Z$ in the following) and $N_c$ states per site (called channels and labelled by $\sigma$). We use the global coordinate index $m=Z (n-1)+ j \sim (n,j)$, which as well contains the unit cell index $n$. 

In particular, we focus on the nearest-neighbour hopping models 
\begin{align}
    \hat{H}_0 =& - \sum_{m=-\infty}^{\infty} \sum_{\sigma, \sigma'=1}^{N_c} (t_{m, \sigma \sigma'} \,\,  |m+1, \sigma\rangle \langle m, \sigma' | + h.c.)
    \\
    &+\sum_{m=-\infty}^{\infty} \sum_{\sigma, \sigma'=1}^{N_c} v_{m,\sigma \sigma'} \,\, |m, \sigma \rangle \langle m, \sigma'|, \\
    t_m &= t_{m+Z}, \quad v_m =v_m^{\dagger} = v_{m+Z},
\end{align}
where $t_m$ and $v_m$ are hopping and potential matrices, respectively, both of the size $N_c \times N_c$.

Solutions $| \psi_{k \alpha}^{(0)} \rangle  $ of the eigenvalue problem $\hat{H}_0| \psi_{k \alpha}^{(0)}\rangle = \epsilon_{k\alpha}^{(0)} |\psi_{k \alpha}^{(0)}\rangle $ are labelled by the Bloch momentum $k$, $ -\pi \leq k < \pi$,  and the band index $\alpha$, $1 \leq \alpha  \leq N_c Z$.
The Bloch Hamiltonian 
\begin{align}
    h_k = \left( \begin{array}{cccccc}
         v_1 & -t_1^{\dagger} &0 & \ldots & 0 & -t_Z e^{-i k} \\
         -t_1 & v_2 & -t_2^{\dagger} &\ldots & 0 & 0\\
         0& -t_2 & & & \vdots & \vdots \\
         \vdots & \vdots & & & - t_{Z-2}^{\dagger} & 0\\
         0 & 0 & \ldots & - t_{Z-2} & v_{Z-1} &-t_{Z-1}^{\dagger} \\
         -t_Z^{\dagger} e^{i k} & 0& \ldots & 0 & -t_{Z-1} & v_Z
    \end{array}\right)
    \label{bloch_hk}
\end{align}
is related to $\hat{H}_0$ via 
\begin{align}
    \hat{H}_0 = \int_{-\pi}^{\pi} dk \,\, | k \rangle \langle k | \otimes h_k, \quad \langle n | k \rangle = \frac{e^{i kn}}{\sqrt{2 \pi}},
\end{align}
with $h_k = \sum_{jj'\sigma\sigma'}(h_k)_{j\sigma,j'\sigma'}|j,\sigma\rangle\langle j',\sigma'|$.
The normalized eigenstates obeying $h_k |\chi_{k\alpha} \rangle = \epsilon_{k\alpha}^{(0)} |\chi_{k\alpha} \rangle$, $  \langle \chi_{k\alpha} |\chi_{k\alpha'} \rangle = \delta_{\alpha \alpha'}$, help us express
\begin{align}
    \langle m,\sigma | \psi_{k \alpha}^{(0)} \rangle = \frac{e^{ikn}}{\sqrt{2\pi}} \langle j,\sigma | \chi_{k \alpha} \rangle .
\end{align}
We note the completeness relation and the identity resolution
\begin{align}
\hat{1} = \sum_{\alpha} \int_{-\pi}^{\pi} dk \,\, |\psi_{k\alpha}^{(0)} \rangle \langle \psi_{k \alpha}^{(0)} |= \sum_{m,\sigma} | m , \sigma \rangle \langle m, \sigma |.
\label{compl_rel_bulk}
\end{align}
Using the notation $\psi_{k\alpha}^{(0)} (m,\sigma) = \langle m, \sigma   |\psi_{k\alpha}^{(0)} \rangle$, we also quote the component-wise form of \eqref{compl_rel_bulk}
\begin{align}
\delta_{mm'} \delta_{\sigma \sigma'} = \sum_{\alpha} \int_{-\pi}^{\pi} dk \,\, \psi_{k\alpha}^{(0) } (m,\sigma) \,\, \psi_{k \alpha}^{(0)\, *} (m' , \sigma').
\label{compl_rel_bulk_matr}
\end{align}

The bulk retarded Green's function $\hat{G}^{(0)} (\omega) = \frac{1}{\omega -\hat{H}_0 + i \eta}$ has the matrix expression
\begin{align}
\nonumber
G^{(0)}_{m\sigma,m'\sigma'} &= \langle m, \sigma |\hat{G}^{(0)} (\omega)| m', \sigma'\rangle \\
\label{bulkG}
&=\int_{-\pi}^{\pi} \frac{d k}{2 \pi} \,\, \langle j, \sigma |\frac{e^{i k (n-n')}}{\omega +i \eta -h_k} | j' , \sigma' \rangle .
\end{align}
In the following we make use of the reduced notations
\begin{align}
G^{(0)}_{m,m'} =\int_{-\pi}^{\pi} \frac{d k}{2 \pi} \,\, \langle j |\frac{e^{i k (n-n')}}{\omega +i \eta -h_k} | j' \rangle \equiv G^{(0)}_{j,j'} (n-n'),
\end{align}
implying that each element $[(\omega + i \eta - h_k)^{-1}]_{jj'}$ is a $N_c\times N_c$ matrix block with internal indices $\sigma,\sigma'$.

Adding an arbitrary potential $\hat{V}$ to $\hat{H}_0$, we break the translational invariance of the 1D lattice model. The retarded Green's function $\hat{G} (\omega)=\frac{1}{\omega + i \eta - \hat{H}}$ of the perturbed model $\hat{H}=\hat{H}_0 + \hat{V}$ satisfies the Dyson equation
\begin{align}
    \hat{G} (\omega)=\hat{G}^{(0)} (\omega) + \hat{G}^{(0)} (\omega) \hat{V}\hat{G} (\omega) .
    \label{dyson_eq}
\end{align}

To mimic a boundary model which is defined in the right half-space ($m \geq 1$), we can choose an infinitely high potential for $m \leq 0$, which would block an occupation of sites in the left half-space. In models which only allow for nearest neighbour hopping it is however sufficient to put a high potential just on the single site $m=0$: It will play the role of the impenetrable barrier between the right and left half-spaces. In the following we restrict ourselves to this class of models and choose $\langle m, \sigma |\hat{V}| m' , \sigma' \rangle = V_0 \,\, \delta_{\sigma \sigma'} \,  \delta_{m,0} \, \delta_{m',0}$, aiming to perform the limit $V_0 \to \infty$ afterwards. A case of longer ranged hoppings will be discussed elsewhere [\onlinecite{Pias_new}]. 

Choosing the ultra-local potential of the above stated form has the advantage that this choice allows us to study the left boundary model with $m \leq -1$ in the same setting. The Dyson equation \eqref{dyson_eq} acquires then the special form
\begin{align}
G_{m \sigma, m' \sigma'}   = G_{m \sigma, m' \sigma'}^{(0)}     +  G_{m \sigma, 0 \sigma_1}^{(0)}  V_0 \,\, G_{0 \sigma_1, m' \sigma'},
\label{GDy}
\end{align}
where we omit for brevity the $\omega$-dependence and implicitly assume a summation over the repeated index $\sigma_1$ (this convention is also used in the following). In the matrix notation, Eq.~\eqref{GDy} is equivalent to
\begin{align}
G_{m, m'} (\omega)  = G_{m, m'}^{(0)} (\omega)    +  V_0 \,\, G_{m,0}^{(0)} (\omega) \,  G_{0 , m'} (\omega).
\label{GDy_matr}
\end{align}

To solve this equation for $G_{m, m'} (\omega)$, we first set $m=0$ and find
\begin{align}
G_{0, m'} (\omega)  = [1-V_0 \, G_{0,0}^{(0)} (\omega)]^{-1} \, G_{0, m'}^{(0)} (\omega) .
\end{align}
Inserting this result back into \eqref{GDy_matr} yields the expression for $\hat{G} (\omega)$ in terms of $\hat{G}^{(0)} (\omega)$:
\begin{align}
G_{m, m'} (\omega)  &= G_{m, m'}^{(0)} (\omega)   \nonumber \\
&+  V_0 \, G_{m,0}^{(0)} (\omega) [1-V_0 \, G_{0,0}^{(0)} (\omega)]^{-1} \, G_{0, m'}^{(0)} (\omega).
\label{GDy_sol}
\end{align}

Now it is appropriate to take the limit of the infinite barrier height $V_0 \to \infty$. It leads to the so called \textit{boundary} Green's function
\begin{align}
G_{m, m'} (\omega)  &= G_{m, m'}^{(0)} (\omega)   \nonumber \\
&- G_{m,0}^{(0)} (\omega) [ G_{0,0}^{(0)} (\omega)]^{-1} G_{0, m'}^{(0)} (\omega).
\label{BGF_main}
\end{align}

\section{Boundary charge and the associated topological invariant}
\label{sec:boundaryCharge}

In this section  we study an excess charge which is accumulated near the hard-wall boundary of a semi-infinite lattice. In particular, we derive an expression for a change $\Delta Q_B$ of the boundary charge $Q_B$ under the lattice shift by one site towards the wall in terms of a topological invariant $I$. The latter is given by a winding number and therefore takes integer values. Thereby we achieve a multichannel generalization of our earlier single-channel result [\onlinecite{pletyukhov_etal_prb_20}, \onlinecite{Pletyukhov_etal2_2020}]. However,  instead of explicitly  constructing the boundary problem eigenstates, which was feasible in the single-channel consideration, we resort now to the representation in terms of the boundary Green's functions, which was introduced in the previous section.

\subsection{Boundary charge definition}

The solution of the boundary model reads
\begin{align}
    \hat{H} |\psi_s \rangle = \epsilon_s | \psi_s \rangle.
\end{align}
Here, $s$ is a general index to label all eigenstates, and it can have both continuous and discrete domains with extended and localized eigenstates, respectively. We note expressions for the
completeness and the boundary Green's function in the basis $\psi_s (m,\sigma) = \langle m, \sigma | \psi_s \rangle$
\begin{align}
    \delta_{mm'} \delta_{\sigma \sigma'} &= \sumint_s \, \psi_s (m,\sigma) \, \psi_s^* (m', \sigma'), \\
    G_{m \sigma, m' \sigma'} (\omega) &= \sumint_s \, \frac{\psi_s (m,\sigma) \, \psi_s^* (m', \sigma')}{\omega + i \eta - \epsilon_s}.
\end{align}

At zero temperature, the charge density equals
\begin{align}
\rho_m &= \sumint_{s} \sum_{\sigma} \Theta (\mu - \epsilon_s) \, \psi_s^* (m, \sigma) \, \psi_s (m, \sigma) \\
&=  \sum_{\sigma}  \int d \omega \, \Theta (\mu - \omega)  \sumint_{s} \psi_s^* (m, \sigma) \, \psi_s (m, \sigma) \, \delta (\omega - \epsilon_s) \nonumber \\
&=  - \frac{1}{\pi}  \sum_{\sigma}  \int d \omega \, \Theta (\mu - \omega) \, \text{Im}\, G_{m \sigma, m \sigma} (\omega) ,
\end{align}
where $\mu$ is the chemical potential. Inserting the solution \eqref{BGF_main} into the above expression we find
\begin{align}
\rho_m - \rho_m^{(0)} &=  \frac{1}{\pi}  \text{Im}  \int d \omega \Theta (\mu - \omega) \nonumber \\
& \times  \text{tr} \left\{ G_{m,0}^{(0)} (\omega) [ G_{0,0}^{(0)} (\omega)]^{-1} G_{0, m}^{(0)} (\omega) \right\},
\label{eq:dens}
\end{align}
where the trace operation is performed in the channel space, and
\begin{align}
    \rho_m^{(0)} = - \frac{1}{\pi}  \text{Im}  \int d \omega \Theta (\mu - \omega) \text{tr} \left\{ G_{m,m}^{(0)} (\omega) \right\} \equiv  \rho_j^{(0)} 
\end{align}
is the density in the translationally invariant model (and thus in the bulk), which only depends on the site index $j$ within the unit cell.

The boundary charge (in units of electron's charge) is defined as
\begin{align}
    Q_B = \sum_{m=1}^{\infty} (\rho_m - \bar{\rho}^{(0)}) f_m,
\end{align}
where $\bar{\rho}^{(0)} = \frac{1}{Z} \sum_{j=1}^Z \rho_j^{(0)}$ is the unit-cell averaged density in the bulk, and $f_m$ is an envelope function mimicking a charge probe (see [\onlinecite{pletyukhov_etal_prb_20}, \onlinecite{Pletyukhov_etal3_2020}] for details). In particular, $f_m \approx 1$ for $m \lesssim M$, and gradually falls off to zero value on the interval $M \lesssim m \lesssim M+N$, with $M,N \gg 1$. Splitting
\begin{align}
    Q_B &= \sum_{m=1}^{\infty} (\rho_m - \rho_m^{(0)}) f_m \label{eq:QFE_0} \\
    &+\sum_{m=1}^{\infty} (\rho_m^{(0)} - \bar{\rho}^{(0)}) f_m,
    \label{eq:QP_0}
\end{align}
one can show that $Q_B = Q'_B +Q_P$, with
\begin{align}
Q'_B &=  \frac{1}{\pi}  \text{Im}   \int d \omega \,\, \Theta (\mu - \omega)
\nonumber \\
& \times \text{tr} \left\{ [ G_{0,0}^{(0)} (\omega)]^{-1} \sum_{m=1}^{\infty} G_{0, m}^{(0)} (\omega) \, G_{m,0}^{(0)} (\omega)\right\}, \label{eq:QFE} \\
Q_P&= -  \sum_{j=1}^Z \frac{j}{Z} \left( \rho_{j}^{(0)} -\bar{\rho}^{(0)} \right).
\label{eq:QP}
\end{align}

The contribution \eqref{eq:QFE} is obtained from \eqref{eq:QFE_0} by approximating $f_m \approx 1$ (which is justified, since \eqref{eq:dens} decays on the scale of a typical localization length $\xi \ll M$). This contribution arises from the density modulation close to the boundary. It contains contributions from the Friedel density oscillations of extended states as well as integer-valued contributions from edge states whose energies might reside in spectral gaps. In Appendix \ref{app:bc} we derive the expression \eqref{fried_spec} for the corresponding integrand in which the sum over $m$ is performed.

In turn, the contribution \eqref{eq:QP} represents the dipole moment of the unit cell (also called the \textit{polarization charge}). It is induced by the spatial variation of $f_m$, which takes place far away from the boundary (see Ref.~[\onlinecite{pletyukhov_etal_prb_20}] for details of deriving \eqref{eq:QP} from \eqref{eq:QP_0}). For the following analysis it is convenient to express
\begin{align}
    \rho_j^{(0)} =&  -\frac{1}{\pi} \int d \omega \,\, \Theta (\mu - \omega) \nonumber \\
    & \times \text{Im tr} \left\{   \int_{-\pi}^{\pi} \frac{d k}{2 \pi} \,\, \langle j | \frac{1}{\omega + i \eta - h_k} |j \rangle \right\} \\
    =& \sum_{\sigma} \sum_{\alpha=1}^{\nu} \int_{-\pi}^{\pi} \frac{dk}{2 \pi} \,\, |\chi_{k \alpha} (j,\sigma)|^2,
\end{align}
where $\nu$ is a number of the fully occupied bands, which depends on the level of the chemical potential $\mu$. Using the normalization of the states $| \chi_{k \alpha} \rangle$ it is also straightforward to show that 
\begin{align}
\bar{\rho}^{(0)} &= \frac{1}{Z} \sum_{j=1}^Z \sum_{\sigma} \sum_{\alpha=1}^{\nu} \int_{-\pi}^{\pi} \frac{dk}{2 \pi} \,\, |\chi_{k \alpha} (j,\sigma)|^2 = \frac{\nu}{Z}.
\end{align}

\subsection{Topological invariant for boundary charge change under the lattice shift}
\label{sec:shift_model}

We shift the lattice by one site towards the boundary and study $\Delta Q_B = \tilde{Q}_B - Q_B$, where $\tilde{Q}_B$ is the boundary charge in the shifted system. For its expression it is sufficient to replace $t_j \to \tilde{t}_j = t_{j+1}$, $v_j \to \tilde{v}_{j} = v_{j+1}$ for $1 \leq j \leq Z-1$, as well as $t_Z \to \tilde{t}_Z = t_1 $, $v_Z \to \tilde{v}_Z = v_1$.

First we evaluate the change in the polarization charge \eqref{eq:QP}
\begin{align}
    \Delta Q_P &= \tilde{Q}_P - Q_P =  -  \sum_{j=1}^Z \frac{j}{Z} \left( \tilde{\rho}_{j}^{(0)}  -\rho_j^{(0)} \right) \\
    &= -  \sum_{j=1}^{Z-1} \frac{j}{Z} \left( \rho_{j+1}^{(0)}  -\rho_j^{(0)} \right) - \rho_{1}^{(0)}  +\rho_Z^{(0)} \\
    &=   \frac{\nu}{Z}    - \rho_{1}^{(0)}  .
\end{align}

\begin{figure}[b]
    \centering
    \includegraphics[width =\columnwidth]{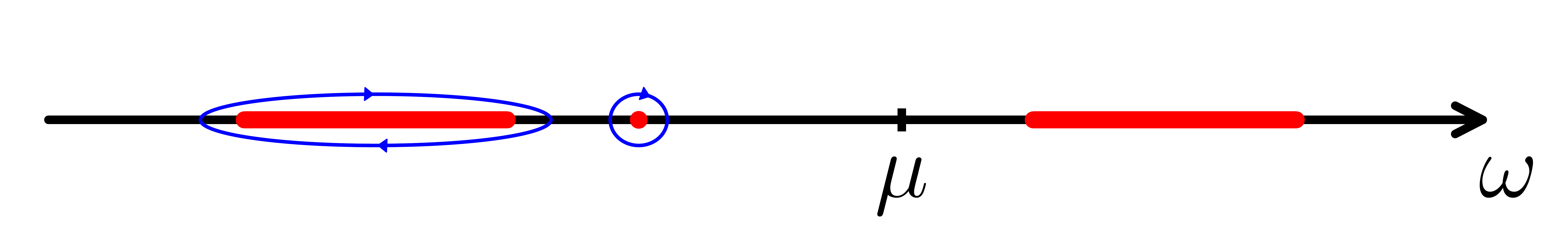}
    \caption{Schematic illustration of the winding of the quantity $\det([G_{0, 0}^{(0)}]^{-1}G_{0, 1}^{(0)}t_{Z})$ in the complex $\omega$ plane. Here, the complex contour traversed  while changing the frequency $\omega$ is indicated by the blue lines, with arrows indicating the direction. In turn, the branch cuts and bound-state poles, which give the spectral ranges of the boundary problem, are indicted as bold red lines and dots, respectively. It is only these non-analytic features (lying below the chemical potential $\mu$) which contribute to the winding number integral.  \label{fig:winding}}
\end{figure}

Defining the quantity
\begin{align}
    I = \Delta Q_B - \frac{\nu}{Z} = \Delta Q'_B - \rho_{1}^{(0)},
    \label{inv_deltaQ}
\end{align}
we show that it can take only integer values and thereby appears to be a topological invariant. In particular, in Appendix \ref{app:shift_QB} we derive the winding number expression
\begin{align}
    I =&  - \frac{1}{\pi} \int d \omega \,\, \Theta (\mu - \omega) \nonumber \\
    & \times \, \text{Im} \, \frac{\partial}{\partial \omega} \, \ln \det \left( [G^{(0)}_{0,0}]^{-1} G_{0,1}^{(0)} t_Z \right).
    \label{wn_I}
\end{align}
As is explained in Sec.~\ref{sec:numerics}, this integral can be interpreted as a winding number of the function $\det \left( [G^{(0)}_{0,0}]^{-1} G_{0,1}^{(0)} t_Z\right)$, as the complex frequency $\omega$ encircles the occupied part of the spectrum, as indicated in Fig.~\ref{fig:winding}. Contributions to the winding number stem from branch cuts (bands) and from poles (edge states).

The result \eqref{inv_deltaQ} is a multichannel generalization of the analogous invariant derived in [\onlinecite{pletyukhov_etal_prb_20}, \onlinecite{Pletyukhov_etal2_2020}] for single-channel models. In Appendix \ref{app:single_ch_boundary} we demonstrate the equivalence of the representation \eqref{wn_I} with those quoted for $I$ in the above cited papers.

\section{Interface charge}
\label{sec:interfaceCharge}

In this section we study an excess charge, which is accumulated at the interface of two semi-infinite models. It is generally defined by 
\begin{align}
    Q_I &= \sum_{m=-\infty}^{\infty} (\rho_m - \bar{\rho}^{(0)}) f_m 
    \label{QI_def} \\
    & = \sum_{m=-\infty}^{\infty} (\rho_m - \rho_m^{(0)}) f_m \label{QI_bare}  \\
    &+\sum_{m=-\infty}^{\infty} (\rho_m^{(0)} - \bar{\rho}^{(0)}) f_m .
    \label{QI_dipole}
\end{align}
Here for negative $m$ the envelope function is defined as a mirror image of its positive part, and in addition we introduce a splitting into the two sums which is analogous to the splitting made in Eqs.~\eqref{eq:QFE_0}, \eqref{eq:QP_0}.

Starting from the translationally invariant model, we consider two types of interfaces: 1) adding a finite potential barrier $V_0$ on site $m=0$ (Sec. \ref{sec:pot_imp}) and 2) weakening the link between sites $m=0$ and $m=1$ by $0\leq\lambda\leq 1$ (Sec. \ref{sec:weak_link}). Similar models have been recently studied in Ref.~[\onlinecite{rhim_2018}] in the search of a unified bulk-boundary correspondence for band insulators.

Since in the underlying model unit cells of the right and left semi-infinite lattices perfectly match, there is no dipole moment in the overall charge distribution, and therefore the contribution \eqref{QI_dipole} identically vanishes. The net interface charge is thus created by removing electrons from the valence band by emerging interface localized states. It is entirely given by the contribution \eqref{QI_bare}, which is alternatively represented by
\begin{align}
    Q_I =&  - \frac{1}{\pi} \, \text{Im}  \int d \omega \Theta (\mu - \omega) \nonumber \\ 
    & \times \sum_{m=-\infty}^{\infty}  \text{tr}  \left\{ G_{m,m} - G_{m,m}^{(0)} \right\}  ,
    \label{QI_green}
\end{align}
where $G$ is the Green's function of the interface model, and $G^{(0)}$ corresponds to the bulk model.

For the above physical reason $Q_I$ is expected to be quantized in integer units of the electron's charge. In the following subsections we derive winding number expressions for $Q_I$, which provide a quantitative confirmation of our expectations. In other words, we analytically prove --- now in the multichannel setting --- the nearsightedness principle postulated in Ref.~[\onlinecite{Kohn_1996}] and used in Ref.~[\onlinecite{Pletyukhov_etal3_2020}] to prove the quantization of the invariant. Details of our intermediate evaluation are provided in Appendix \ref{app:cut_link}.

\subsection{Potential impurity}
\label{sec:pot_imp}

In this subsection we consider a model characterized by the Green's function \eqref{GDy_sol}, that is featuring the additional impurity potential $V_0$ at the site $m=0$. On the basis of \eqref{QI_green} it holds
\begin{align}
    Q_I =&  - \frac{1}{\pi} \, \text{Im}  \int d \omega \,\, \Theta (\mu - \omega) \nonumber \\ 
    & \times V_0 \,\,   \text{tr}  \left\{ [1-V_0 \, G_{0,0}^{(0)} ]^{-1} \sum_{m=-\infty}^{\infty} G_{0, m}^{(0)} \, G_{m,0}^{(0)} \right\} .
    \label{QI_V0}
\end{align} 
Making use of \eqref{Gm_sum}, we express 
\begin{align}
    Q_I =&  - \frac{1}{\pi} \, \text{Im}  \int d \omega \,\, \Theta (\mu - \omega) \nonumber \\ 
    & \times   \text{tr}  \left\{    (1-V_0 F_0 )^{-1} \, \frac{\partial (1-V_0 F_0)}{\partial \omega} \right\} \\
    =& - \frac{1}{\pi}  \int d \omega \,\, \Theta (\mu - \omega) \nonumber \\ 
    & \times  \, \text{Im} \, \frac{\partial}{\partial \omega} \ln \det (1-V_0 \, G_{0,0}^{(0)} ),
    \label{QI_V0_wn}
\end{align}
where $F_0 = G_{0,0}^{(0)}$ [cf. Eq.~\eqref{eq:def_F0}]. Like $I$ in \eqref{wn_I}, $Q_I$ acquires integer values. They are equal to winding numbers of the function $\det (1-V_0 \, G_{0,0}^{(0)} )$ in the complex $\omega$-plane [cf. Fig.~\ref{fig:contour}].

At $V_0=0$ we restore the translational invariance, and therefore $Q_I (V_0 =0)=0$.

In the limit $V_0 \to \infty$ (two isolated subsystems excluding site $m=0$) we obtain
\begin{align}
Q_I (V_0 \to \infty )=&- \frac{1}{\pi}  \int d \omega \,\, \Theta (\mu - \omega)  \, \text{Im} \, \frac{\partial}{\partial \omega} \ln \det G_{0,0}^{(0)} .
    \label{QI_V0_limit}
\end{align}

\subsection{Link weakening}
\label{sec:weak_link}

In this subsection we generalize the result of Ref.~[\onlinecite{Pletyukhov_etal3_2020}] (see Appendix C therein) to the multichannel case.

We add the perturbation $V_{\lambda} = (1-\lambda)\,\, [t_Z^{\dagger} \,|0 \rangle  \langle 1| + t_Z \, |1 \rangle  \langle 0|]$, with $0 \leq \lambda \leq 1$,  to the translationally invariant model. The corresponding Dyson equation reads
\begin{align}
G_{m,m'} = G_{m,m'}^{(0)} +  G_{m,1}^{(0)} \, t_{\lambda} \, G_{0,m'} + G_{m,0}^{(0)} \, t_{\lambda}^{\dagger} \, G_{1,m'}  .
\label{dyson_link}
\end{align}
where $t_{\lambda}= (1-\lambda)  t_Z $. It follows
\begin{align}
G_{0,m'} &= G_{0,m'}^{(0)} +  G_{0,1}^{(0)} \, t_{\lambda} \, G_{0,m'} +     G_{0,0}^{(0)} \, t_{\lambda}^{\dagger} \, G_{1,m'} , \label{g0m} \\
G_{1,m'} &= G_{1,m'}^{(0)} +   G_{1,1}^{(0)} \, t_{\lambda} \, G_{0,m'} +  G_{1,0}^{(0)} \, t_{\lambda}^{\dagger} \, G_{1,m'} . \label{g1m}
\end{align}
Solving these equations for $G_{0,m'}$ and $G_{1,m'}$ (see Appendix \ref{app:cut_link}) and inserting the obtained solutions into \eqref{dyson_link}, we evaluate
\eqref{QI_green}. Performing the sum over all sites, we derive the winding number expression
\begin{align}
Q_I =&  - \frac{1}{\pi} \int d \omega \,\, \Theta (\mu - \omega)  \nonumber \\
& \times \text{Im}  \, \frac{\partial}{\partial \omega}  \ln \det \left[ (1- \lambda^2) \left( \Lambda + \frac12 \right) +\lambda^2 \right], \label{QI_lambda} 
\end{align}
where $\Lambda = \frac12 - F_0^{-1} \, G_{0,1}^{(0)} \, t_Z \, F_0$ [cf. Eq.~(\ref{eq:def_Lambda})].

At $\lambda=1$ we restore the translational invariance, and therefore $Q_I (\lambda=1)=0$.

At $\lambda=0$ (two isolated subsystems including site $m=0$) we obtain
\begin{align}
Q_I (\lambda=0) &=  - \frac{1}{\pi} \int d \omega \,\, \Theta (\mu - \omega) \nonumber \\
& \times \text{Im}  \, \frac{\partial}{\partial \omega} \,  \ln \det \left( \Lambda + \frac12 \right)  . \label{QI_lambda0} 
\end{align}
Using Eqs.~\eqref{LL_exch} and \eqref{G01expr}, we deduce
\begin{equation}
    \label{eq:linkWeakeningAlternative}
    \det \left[ (1- \lambda^2) \left( \Lambda + \frac12 \right) +\lambda^2 \right] = \det \left[1 - \left(1-\lambda^2\right) \, G_{0,1}^{(0)} t_Z\right],
\end{equation}
which is a convenient representation for evaluating the invariant \eqref{QI_lambda} at finite $\lambda$.

In Appendix \ref{app:single_ch_interface} we demonstrate how  to reproduce from \eqref{QI_lambda} the single-channel expression for the interface charge previously derived in Ref.~[\onlinecite{Pletyukhov_etal3_2020}].

\section{Numerical Results}
\label{sec:numerics}

In this section we demonstrate the validity of the above defined novel topological invariants, particularly Eqs.~(\ref{wn_I}), (\ref{QI_V0_wn}), and (\ref{QI_lambda}). To this end we evaluate these winding numbers for some randomly generated multichannel models and compare their values to the corresponding boundary and interface charges, evaluated from their original definitions. As it turns out, even for the minimal nontrivial (that is, allowing for the "spin-orbit" coupling) multichannel models ($N_c=Z=2$), the analytical evaluation of these winding numbers is unfeasible, so we have to resort to a numerical calculation.

This section is organzied as follows: in Sec. \ref{sec:howToWinding} we describe our strategy of the numerical evaluation of the winding numbers. In particular, we describe an algorithm which allows for an efficient computation of these invariants. In Sections \ref{sec:numBoundary} and \ref{sec:numInterface} we show representative data for the winding numbers corresponding to the boundary and interface charges, respectively.

\subsection{Numerical implementation of the winding numbers}
\label{sec:howToWinding}

All three winding numbers discussed in this work are defined in terms of bulk Green's functions, which in turn are defined via a quasimomentum integral, see Eq.~(\ref{bulkG}). In the numerical evaluation, we approximate this integral by a sum over $N_k$ evenly spaced (by $\delta_k = \frac{2 \pi}{N_k}$) momenta $k_l = -\pi + l\delta_k$, with $l=0,\dots,N_k-1$, i.e.~
\begin{equation}
    \label{GF_approx}
    G^{(0)}_{m\sigma, m^{\prime}\sigma^{\prime}} = \lim_{N_k \to \infty} \frac{\delta_k}{2\pi} \sum_{k_l} \left< j, \sigma\left|\frac{e^{ik_l(n-n^{\prime})}}{\omega + i\eta - h_{k_l}}\right|j^{\prime},\sigma^{\prime}\right>.
\end{equation}

\begin{figure*}
    \includegraphics[scale=0.9]{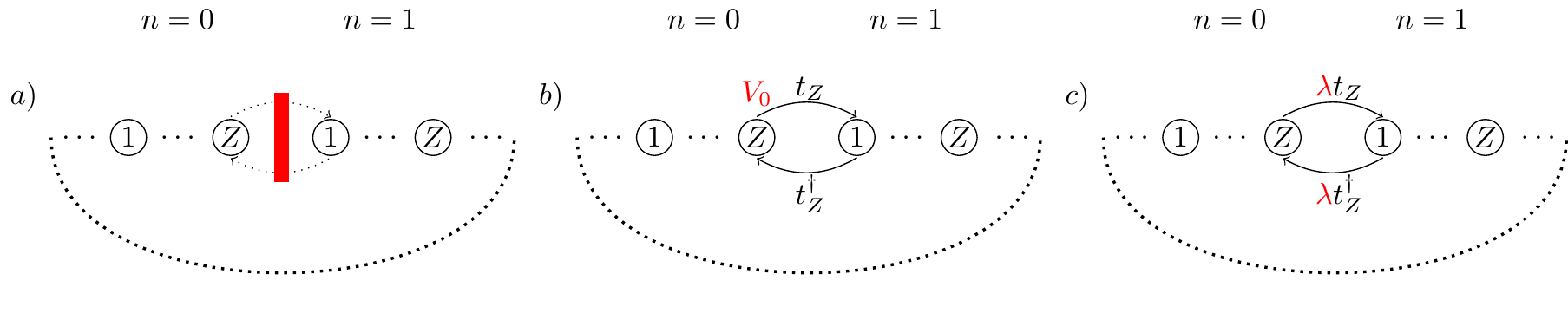}
    \caption{Diagram depicting the various boundary configurations considered here: hard wall (a), potential impurity (b) and weakened link (c). In all three cases we consider a closed ring of $N_Z$ unit cells, as indicated by the bent dotted lines underneath the diagrams which contain $N_Z-2$ additional unit cells respectively. The labels above the diagrams correspond to the numbering of the unit cells as discussed in Sec.~\ref{sec:BGF}. The feature distinguishing a given boundary configuration from the translationally invariant ring is highlighted in red, respectively. a) shows the boundary configuration of a hard wall, which is used in the computation of the shift in boundary charge. The link between unit cells $n=0$ and $n=1$ is cut completely and the boundary charge is measured to the right, i.e.~starting with unit cell $n=1$. The envelope function falls off to zero to the right of $n=1$, around $n \sim N_Z/4$. b) shows the potential impurity boundary configuration, where the onsite potential $V_0$ is added onto site $m=0$ ($n=0,j=Z$) of the translationally invariant ring. Note that in the limit $V_0\to \infty$ we effectively obtain a hard wall again (as in a)), but with the site $m=0$ missing. This boundary configuration is used to evaluate the interface charge, that is the envelope function extends across the impurity to the left and right, over the range between $-N_Z/4$ and $N_Z/4$. c) shows the boundary configuration of a weakened link, where the hopping matrices between unit cells $n=0$ and $n=1$ are reduced by a factor $0\leq\lambda\leq1$. For $\lambda=1$ the translationally invariant ring is recovered, the limit $\lambda=0$ corresponds to the hard wall configuration of a). The interface charge is again computed using an envelope function extending across the impurity to the left and right.}
    \label{fig:boundaryConfig}
\end{figure*}

The various winding numbers computed from these Green's functions are then compared to exact diagonalization data computed from some finite size Hamiltonian, suitably defined according to the boundary configuration in question, i.e. \textit{hard wall}, \textit{potential impurity} or \textit{link weakening} [see Fig.~\ref{fig:boundaryConfig}]. These finite size Hamiltonians have some dimension $D = N_{\text{sites}} N_c$, where $N_{\text{sites}}$ denotes the number of sites (chosen in the following to accommodate an integer number $N_Z$ of unit cells, that is $N_{\text{sites}}=Z N_Z$), and are characterized by some spectrum $\epsilon_s$ (where $s=1,\dots,D$) and bandwidth $B=\max_s \epsilon_s - \min_s \epsilon_s$. We define the average level spacing of a finite size Hamiltonian as
\begin{equation}
    \delta_{\epsilon} = \frac{B}{D}.
\end{equation}
The Green's function evaluated via the sum over momenta converges to that of an infinite system, if
\begin{equation}
    \label{momentumSpacing}
    \sup_k || \delta_k \partial_k h_k|| \ll \delta_{\epsilon}.
\end{equation}
 Here, $||\dots||$ denotes any matrix norm. In the following we will thus assume that all Green's functions are computed according to Eq.~(\ref{GF_approx}) where the momentum spacing $\delta_k$ is not sent to zero, but kept finite and chosen according to Eq.~(\ref{momentumSpacing}). In practice we choose $\delta_k = \frac1D$ and find that with this choice Eq.~(\ref{momentumSpacing}) is always fulfilled.

\begin{figure}
    \centering
    \includegraphics[width = \columnwidth]{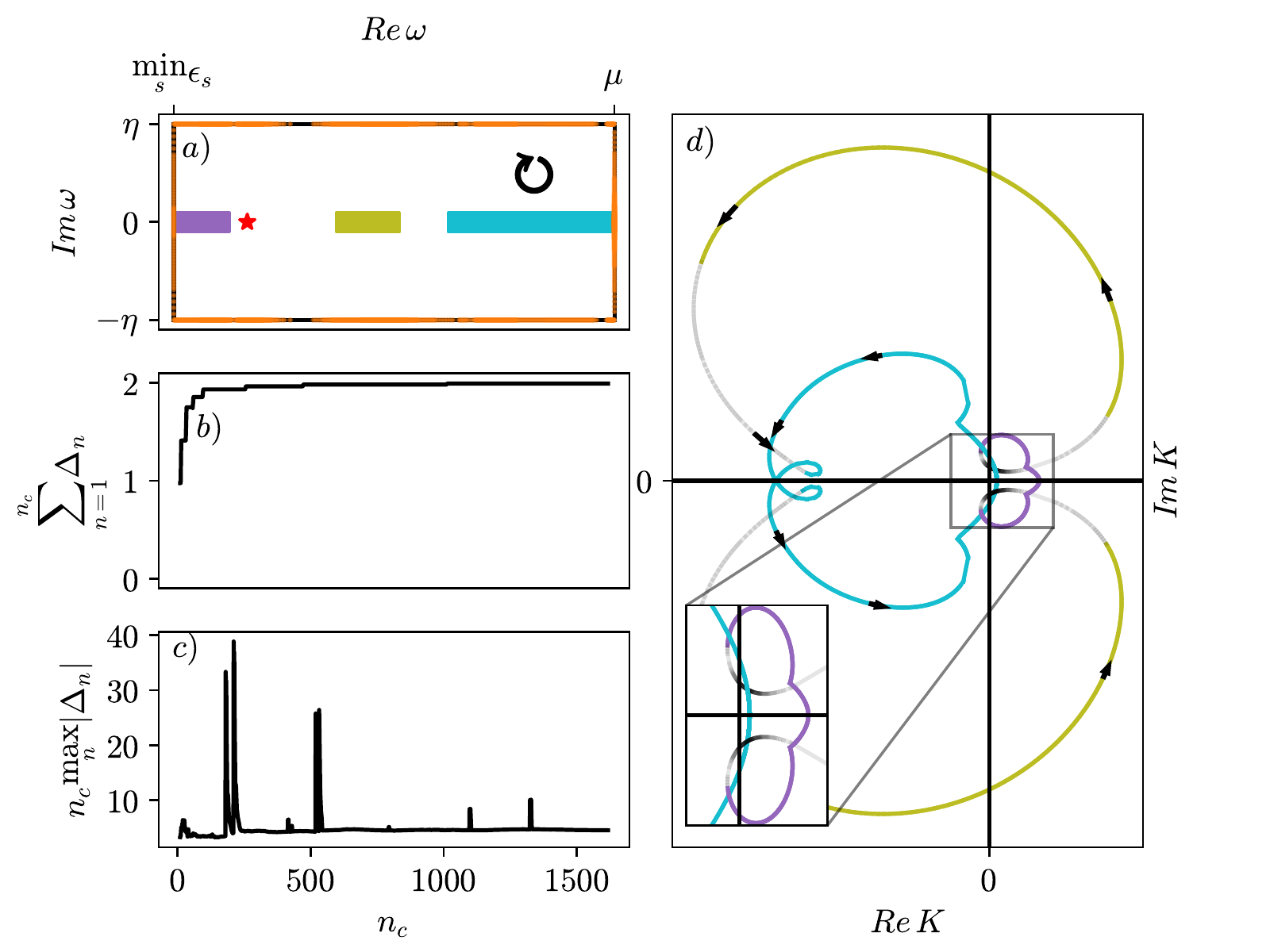}
    \caption{Representative data of the evaluation of a $K=\det \left( [G^{(0)}_{0,0}]^{-1} G_{0,1}^{(0)} t_Z \right)$ winding number for a randomly generated $N_c=Z=3$ model with three occupied bands (components of the Hamiltonian are listed in the Supplemental Material [\onlinecite{supp_mat}]). Panel a) shows samples comprising the energy contour (in orange) and the distribution of bands (colored) and gaps (white) and an edge state pole marked by the red star. The solid black rectangle underneath the samples denotes the complete contour, the curved arrow in the upper right indicates its orientation. Panels b) and c) show the estimate of the winding number $\sum_{n=1}^{n_c} \Delta_n$ and the scaled error $n_c\max_n|\Delta_n|$ as functions of the number of samples $n_c$, respectively. Panel d) shows the $K-$contour with arrows indicating the orientation. It is characterized by a winding number equal to two, which accords with the value obtained in panel b). The colors of the various sections of the $K-$contour correspond to the bands, while gaps are denoted by gray color.}
    \label{fig:contour}
\end{figure}

Having settled the question of how to compute the bulk Green's functions we now turn to the question of how to evaluate the winding numbers themselves. All three winding numbers defined in this work are given in terms of energy integrals of the following kind (up to overall minus signs)
\begin{equation}
    \text{wn}\left[K\right] = \int \frac{d \omega}{\pi} \Theta(\mu - \omega) \text{Im} \, \partial_{\omega} \ln K(\omega),
\end{equation}
where 
\begin{equation}
    \label{eq:KChoices}
    K \in \begin{cases}
    \det \left( [G^{(0)}_{0,0}]^{-1} G_{0,1}^{(0)} t_Z \right) \\
    \det \left(1-V_0 G_{0,0}^{(0)} \right) \\
    \det \left[ 1 - \left(1-\lambda^2\right) G_{0,1}^{(0)} t_Z \right]
    \end{cases},
\end{equation}
depending on the quantity in question. Note that in all three cases $K^{\ast} = K\rvert_{\omega + i\eta \to \omega - i \eta}$. The winding number expression can thus be recast as follows:
\begin{align}
    \text{wn}\left[K\right] 
    &= \frac{1}{2\pi i} \left[\int\displaylimits_{-\infty}^{\mu} \partial_{\omega} \ln K\rvert_{\omega + i\eta} + \int\displaylimits_{\mu}^{-\infty} \partial_{\omega} \ln K\rvert_{\omega - i\eta}\right]\\
    &= \oint_{\mathcal{C}} \frac{d \omega}{2\pi i} \partial_\omega \ln K(\omega)
    = \lim_{n_c \to \infty}\sum_{n=1}^{n_c} \Delta_n,
\end{align}
where $\mathcal{C}$ denotes a rectangular contour with infinitesimal width $2\eta$ in the imaginary direction, ranging (in the real direction) from the lowest band edge $\min_s \epsilon_s$ (instead of $-\infty$) to the chemical potential $\mu$, and where 
\begin{equation}
    \Delta_n =\begin{cases} \frac{\arg K(\omega_{n+1}) - \arg K(\omega_n)}{2\pi} \quad \text{if} \quad n + 1 \leq n_c\\
    \frac{\arg K(\omega_1) - \arg K(\omega_{n_c})}{2\pi} \quad \text{if} \quad n = n_c
    \end{cases},
\end{equation}
with the $n_c$ ordered samples $\left\{\omega_n\right\}_{n=1,\dots,n_c}$ from the contour $\mathcal{C}$ (see panel a) of Fig.~\ref{fig:contour}). 

Note that the value of the integral does not change upon a continuous deformation of the rectangular contour $\mathcal{C}$, as long as one does not cross any non-analytic features of $K$ (e.g.~poles or branch-cuts) in the process. Due to the symmetry under complex conjugation mentioned above it is however numerically advantageous to resort to a rectangular contour, since any evaluation of $K$ automatically yields two samples needed for the computation of the winding number (at $\omega$ and at $\omega^{\ast}$). In the following we will thus assume that we always exploit this symmetry.

Since the evaluation of $K$ for a single sample $\omega_n$ involves the computationally expensive sum over $N_k$ momenta mentioned above, it is desirable to limit the number of samples $n_c$ to a minimum. We achieve this by not using a fixed grid of evenly spaced samples along the energy contour, but by starting with a small number of evenly spaced samples (at least three), and then iteratively adding samples in between those points, for which the absolute value of the phase difference $|\Delta_n|$ is largest. The algorithm consists of the following steps:
\begin{itemize}
    \item Choose $N_{\text{start}}$ evenly spaced samples $\omega_n$ on the energy contour $\mathcal{C}$, compute the corresponding values of $K(\omega_n)$, and with those the phase differences $\Delta_n$. At each stage of the algorithm, the sum $\sum_{n=1}^{n_c} \Delta_n$ provides an estimate of the value of the winding number and $\max_n |\Delta_n|$ provides a measure for the error in the winding number, which is expected to behave like $\sim \mathcal{O}(n_c^{-1})$.
    \item As long as there are neighbouring points on the $K$-contour, for which the absolute value of the phase difference $|\Delta_n|$ is larger than some threshold $\delta$: Determine $n_0$, such that $|\Delta_{n_0}| = \max_n |\Delta_n|$, evaluate $K$ at the value $\omega_{\text{new}}$ on the energy contour which lies in between (along the energy contour) $\omega_{n_0}$ and $\omega_{n_0 + 1}$. Compute the phase differences for the two new pairs of neighbours ($\left\{\omega_{n_0},\omega_{\text{new}}\right\}$ and $\left\{\omega_{\text{new}},\omega_{n_0+1}\right\}$) which emerge due to the addition of this new sample, recompute the sum over all phase differences. 
\end{itemize}
Once the absolute values of all phase differences are below a certain threshold, the (rounded) final estimate $\sum_{n=1}^{n_c} \Delta_n$ is taken as the result of the winding number computation. Note that whenever we evaluate $K$ at a given $\omega$, we also get the value at $\omega^{\ast}$ \textit{for free}, by exploiting the above-mentioned symmetry.

Fig.~\ref{fig:contour} shows an example of the computation of a winding number for $K = \det \left( [G^{(0)}_{0,0}]^{-1} G_{0,1}^{(0)} t_Z \right)$, $N_c=Z=3$, three occupied bands and a randomly generated Hamiltonian whose matrix elements are listed in the Supplemental Material [\onlinecite{supp_mat}]. 

Panel a) of Fig.~\ref{fig:contour} shows the energy contour with samples $\omega_n$ (in orange) and the extent of the bands as indicated by the colored regions (white areas correspond to gaps, and the red star denotes the pole). In practice, the extent $\eta$ of the contour in the imaginary direction is not infinitesimally small, but rather finite and small compared to the bandwidth. In this plot, and in what follows, we always use $\eta = 10\delta_{\epsilon}$. One can clearly see that the density of samples along the contour varies greatly, with densely populated, but also completely blank sections. Together with the quick convergence of the winding number discussed below, this demonstrates the usefulness of the above described algorithm, as compared to a fixed, uniform grid, which would lead to many unnecessary evaluations of $K$. 

Panels b) and c) show the estimate of the winding number $\sum_{n=1}^{n_c} \Delta_n$ and the scaled error $ n_c \max_n |\Delta_n|$ as functions of the number of samples $n_c$, respectively. It is apparent that convergence is reached quickly and that the error falls off as $\sim\mathcal{O}(n_c^{-1})$, as expected. Occasional peaks in the scaled error correspond to the \textit{discovery} of new parts of the $K$-contour by the algorithm. Once such a new section has been discovered it is quickly \textit{smoothed out} by considering more samples in the corresponding region of the energy contour, reducing the scaled error again.

Panel d) shows the contour described by $K(\omega_n)$ in the limit where the value of the winding number has converged. The colors of the various sections of the contour correspond to the equally colored bands in panel a), while sections corresponding to gaps are colored in gray.

In this example (and in all following calculations), we use $N_{\text{start}} = 10$ and $\delta = \frac{1}{360}$, i.e.~in the converged limit no pair of neighbouring points has an absolute phase difference of more than one degree as measured from the origin. Although we have discussed the special case of $K = \det \left( [G^{(0)}_{0,0}]^{-1} G_{0,1}^{(0)} t_Z \right)$ here, we report that other combinations of Green's functions, and in particular the relevant functions defined in Eq.~(\ref{eq:KChoices}), show qualitatively similar behavior as the example shown in Fig.~\ref{fig:contour}.

\subsection{Numerical validation of the boundary charge invariant}
\label{sec:numBoundary}

The boundary charge invariant [Eq.~(\ref{wn_I})] is related to the change in boundary charge under a shift of the lattice by one site towards the boundary. In order to compute the boundary charge we define a finite size Hamiltonian with $N_Z$ unit cells and cut the system, as depicted in Fig.~\ref{fig:boundaryConfig}(a).

In order to evaluate the change in the boundary charge upon continuous shift of the system towards the left boundary we consider the following form for the onsite potentials $v_n$ and hopping matrices $t_n$:
\begin{align}
    \label{eq:boundaryChargeBlocks_v}
    v_n &= v_0 + \cos\left(\varphi + \frac{2\pi (n - 1)}{Z}\right) \delta_v, \\
    \label{eq:boundaryChargeBlocks_t}
    t_n &= t_0 + \cos\left(\varphi + \frac{2\pi (n - 1)}{Z}\right) \delta_t,
\end{align}
where $v_0 = v_0^{\dagger},\delta_v = \delta_v^{\dagger},t_0$ and $\delta_t$ are random $N_c\times N_c$-dimensional matrices. Here, the phase variable $\varphi$ controls the shift of the right subsystem toward the boundary, in particular a shift of $\varphi \to \varphi + \frac{2\pi}{Z}$ shifts it  by one site. 

Note that in the previous work [\onlinecite{pletyukhov_etal_prb_20}] a more general class of phase dependencies for the components of the Hamiltonian was considered, with higher Fourier coefficients and additional random phase shifts. Restricting ourselves to the form stated above however suffices to demonstrate the validity of the novel winding number invariant defined in this work, since no conceptual differences arise in the spectral properties of the Hamiltonian upon including such higher Fourier coefficients.

We diagonalize the finite size Hamiltonian with open boundary conditions and $N_Z=100$ unit cells in order to compute the boundary charge $Q_B(\varphi)$ in the right subsystem  (i.e. between the cut link on the left $n=1$ and the envelope's function fall-off region $n \sim N_Z/4$ on the right) for a given chemical potential $\mu$ and phase $\varphi$. The change in boundary charge upon shift by one site is then simply given by $\Delta Q_B(\varphi) = \tilde{Q}_B(\varphi)- Q_B(\varphi)$ where $\tilde{Q}_B(\varphi) = Q_B\left(\varphi + \frac{2\pi}{Z}\right) $.

\begin{figure}
    \centering
    \includegraphics[width = \columnwidth]{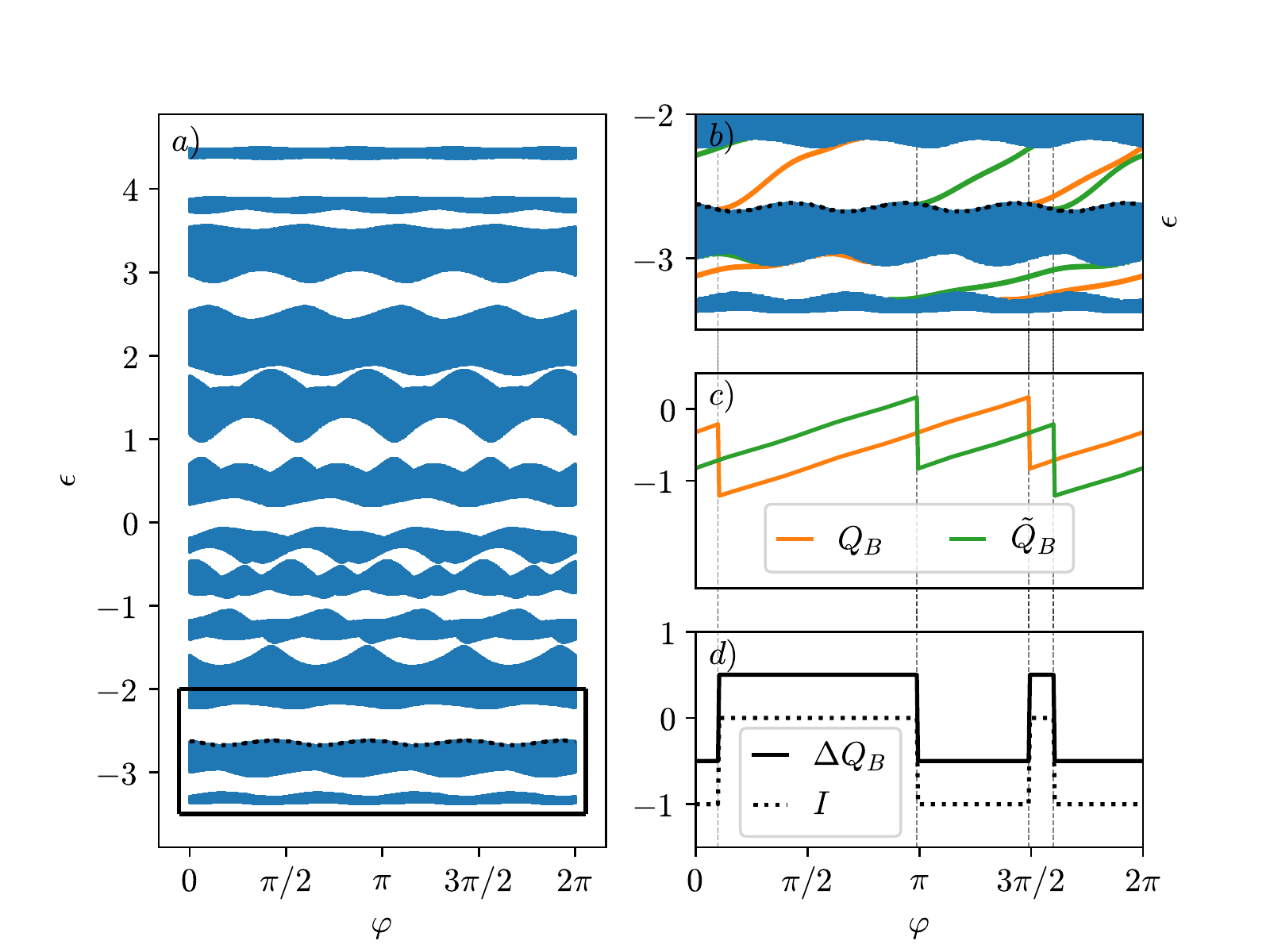}
    \caption{Representative data of the comparison of boundary charge $Q_B$ and boundary charge invariant $I$ for a randomly generated $N_c=3, Z=4$ model with two occupied bands (components of the Hamiltonian are listed in the Supplemental Material [\onlinecite{supp_mat}]). Panel a) shows the extent of the bands and gaps of the model as a function of the phase $\varphi$, the chemical potential (dashed line) and a rectangular box indicating the relevant energy window. Panel b) shows a zoom of this region with additional orange (green) lines indicating edge states localized on the right from the cut link in the system (the shifted system), as computed from the diagonalization of a finite size Hamiltonian with $N_Z=100$ unit cells respectively. The chemical potential is indicated as a dashed line. Panel c) shows both the boundary charge $Q_B$ and shifted boundary charge $\tilde{Q}_B = Q_B\left(\varphi + \frac{2\pi}{Z}\right)$ as functions of $\varphi$. Panel d) shows the change in boundary charge $\Delta Q_B$ and the winding number invariant $I$ [Eq.~(\ref{wn_I})]. The offset between these two curves is exactly $-\frac{\nu}{Z} = -\frac12$ for all values of $\varphi$.}
    \label{fig:boundaryCharge}
\end{figure}

Fig.~\ref{fig:boundaryCharge} shows a comparison of the boundary charge $Q_B$ (computed in the way stated above), the change in boundary charge $\Delta Q_B$ and the winding number invariant $I$, all as functions of the phase $\varphi$ for a randomly generated $N_c=3, Z=4$ model with two occupied bands. The matrices $v_0,\delta_v,t_0,\delta_t$ used in this example are listed in the Supplemental Material [\onlinecite{supp_mat}].

Panel a) of Fig.~\ref{fig:boundaryCharge} shows a band structure of the translationally invariant ring as a function of the phase $\varphi$. Some of the $N_cZ = 12$ bands touch at special phases, but the gap above the second band remains open throughout the whole phase cycle. The dashed line at the top of the second band corresponds to the chemical potential and the rectangular box denotes the relevant region of the band structure.

Panel b) shows a zoom on the relevant bands and gaps (especially the second gap), the chemical potential as a dashed black line and the energies of edge states localized on the right from the cut link in both the initial (orange) and shifted (green) finite system with $N_Z=100$ unit cells, all as functions of the phase $\varphi$.

Panel c) shows the boundary charge $Q_B$ and the boundary charge of the shifted system $\tilde{Q}_B$ as a function of the phase $\varphi$. One can see that the boundary charge behaves almost linearly $\sim \frac{\varphi}{\pi}$ interrupted by two jumps by $-1$ at those phases, where edge states leave the topmost occupied band.

Panel d) shows the difference in boundary charge $\Delta Q_B$ and the winding number $I$ [Eq.~(\ref{wn_I})], both as functions of the phase $\varphi$. One can see that the two lineshapes are identical, up to an offset of $ -\frac{\nu}{Z} = -\frac12$, demonstrating the validity of the boundary charge invariant for this randomly generated model.

We report that in preparation of this work we have simulated numerous of these random models with varying $N_c,Z$ and number of occupied bands and have never seen failure of Eq.~(\ref{wn_I}). We furthermore report that in no case we have seen $I<-N_c$ or $I>0$.

\subsection{Numerical validation of the interface charge invariants}
\label{sec:numInterface}

In addition to the winding number relating to the change in boundary charge discussed above, this work also introduces two winding numbers relating to the interface charge that accumulates around an impurity in an otherwise translationally invariant system. We consider two types of impurities: a local potential on site $m=0$ and a weakened link between sites $m=0$ and $m=1$. 

In the following we discuss one example for each of these two cases respectively, comparing the value of the winding numbers to the interface charges both as functions of the strength of the impurity. To this end we model the translationally invariant system as a ring with $N_Z$ unit cells and add the respective impurity [see Fig.~\ref{fig:boundaryConfig}(b,c)]. We then diagonalize these finite size Hamiltonians and compute the sum of the boundary charges to the left and to the right of the impurity. We denote these two types of boundary charges by $Q_B^{(R)}$ (to the right of the impurity) and $Q_B^{(L)}$ (to the left of the impurity, also including the impurity site $m=0$), such that the interface charge is given by $Q_I^{\text{def}} = Q_B^{(R)} + Q_B^{(L)}$, where we use the label "$\text{def}$" in order to distinguish the interface charge calculated on the basis of the definition \eqref{QI_def} from the winding number invariants, which we label by $Q_I$ [see Eqs.~(\ref{QI_V0_wn}), (\ref{QI_lambda})].

\subsubsection{Potential impurity}

We consider a finite size Hamiltonain with $N_Z=100$ unit cells in a ring configuration and add the potential impurity $V_0$ on site $m=0$, as depicted in Fig.~\ref{fig:boundaryConfig}(b). We diagonalize this Hamiltonian and use the resulting wavefunctions to compute the interface charge.

\begin{figure}
    \centering
    \includegraphics[width = \columnwidth]{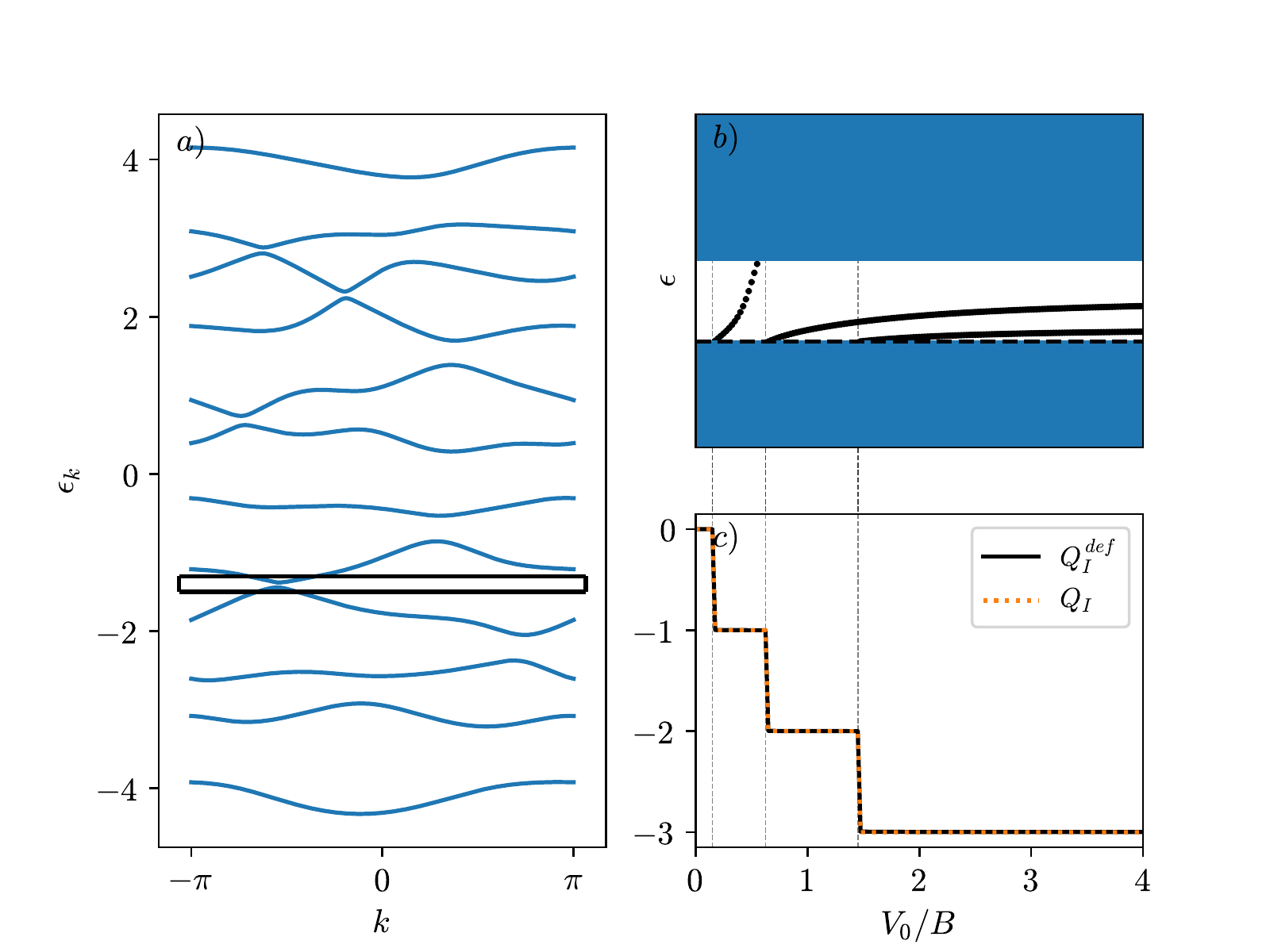}
    \caption{Representative data for the comparison of the interface charge computed on the basis of Eq.~\eqref{QI_def} and the interface charge invariant \eqref{QI_V0_wn} for a randomly generated $N_c = 4, Z=3$ model with a potential impurity of strength $V_0$ on site $m=0$ and four occupied bands. The components of the Hamiltonian are listed in the Supplemental Material [\onlinecite{supp_mat}]. Panel a) shows the band structure of the translationally invariant system ($V_0=0$) as a function of quasimomentum $k$ and a rectangle indicating the relevant energy window above the fourth band. Panel b) shows the gap above the fourth band and the energies of bound states in a $N_Z=100$ model as a function of the impurity strength $V_0$ measured in units of the bandwidth $B$. Panel c) shows the lineshapes of  $Q_I^{\text{def}}$ and  $Q_I$, both as functions of $V_0$, which perfectly overlap.}
    \label{fig:impurityExample}
\end{figure}

Fig.~\ref{fig:impurityExample} shows a comparison of the such computed interface charge (which we denote by $Q_I^{\text{def}}$) and the corresponding topological invariant $Q_I$ (as defined in Eq.~(\ref{QI_V0_wn})) for a randomly generated $N_c=4, Z=3$ model with four occupied bands. The Hamiltonian blocks are listed in the Supplemental Material [\onlinecite{supp_mat}].

Panel Fig.~\ref{fig:impurityExample}(a) shows the band structure of the translationally invariant model (i.e.~for $V_0=0$) versus $k$, with a rectangle indicating the relevant region around the fourth gap. 

Panel b) shows a zoom onto this region with additional lines indicating the energies of bound states as functions of the impurity strength $V_0$ in units of the bandwidth $B$. One can see that as the impurity strength is increased, in total three bound states emerge from the fourth band, two of which remain in the gap for all values of $V_0$, while the third bound state joins the overlying band. 

Panel c) shows the interface charge $Q_I^{\text{def}}$ and the invariant $Q_I$ [Eq.~(\ref{QI_V0_wn})], both as functions of the impurity strength. The lineshapes of $Q_I$ and $Q_I^{\text{def}}$ perfectly overlap, demonstrating the validity of this topological invariant.

Again, we report that in preparation of this work we have tested many random models, always finding agreement of \eqref{QI_def} with Eq.~(\ref{QI_V0_wn}).

\subsubsection{Link weakening}

Next, we consider a finite size Hamiltonain with $N_Z=100$ unit cells in a ring configuration and weaken the link between sites $m=0$ and $m=1$ by $0\leq\lambda\leq 1$, as depicted in Fig.~\ref{fig:boundaryConfig}(c). We diagonalize this Hamiltonian and use the resulting wavefunctions to compute the interface charge.

\begin{figure}
    \centering
    \includegraphics[width = \columnwidth]{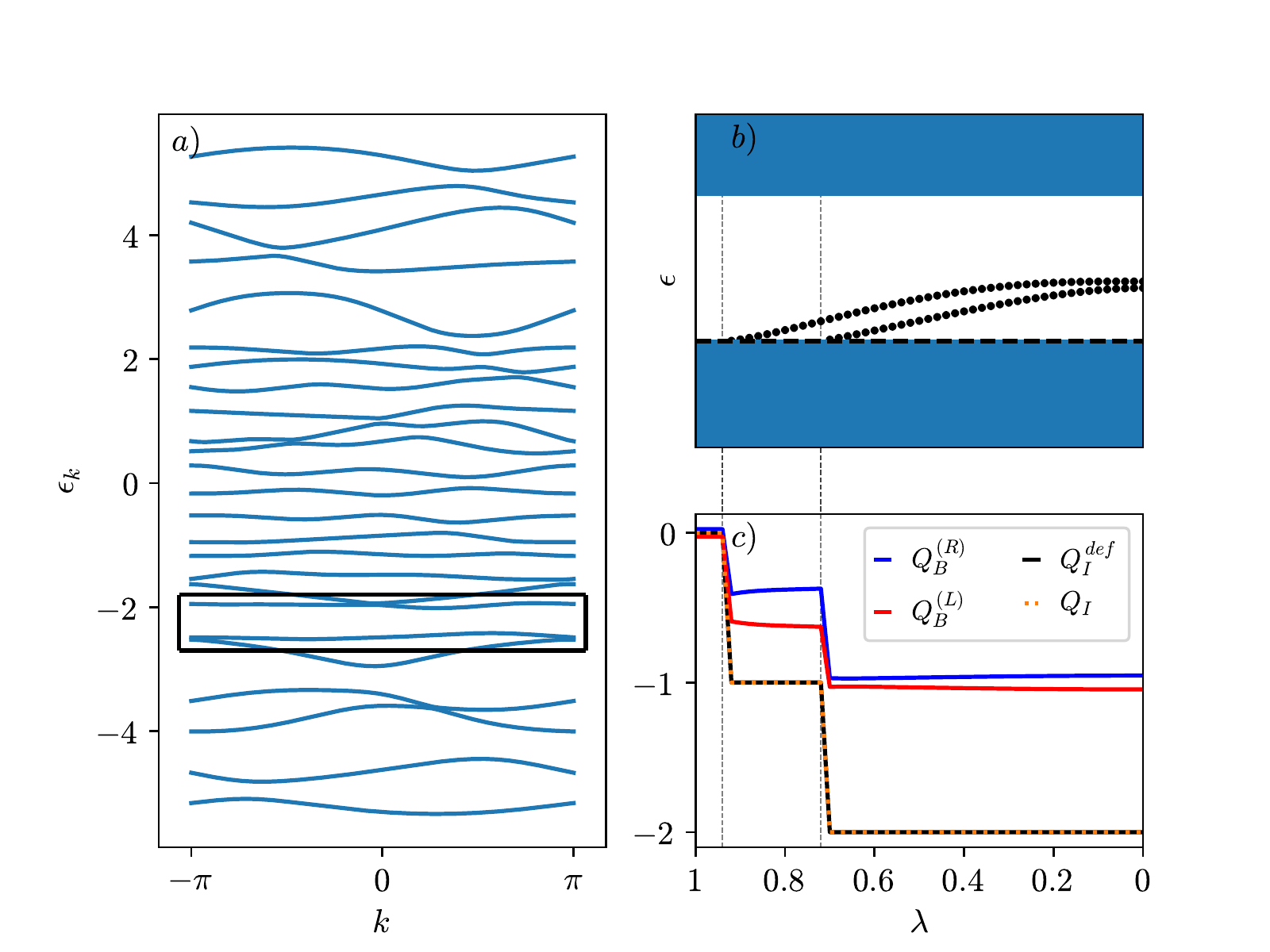}
    \caption{Representative data for the comparison of the interface charge computed on the basis of Eq.~\eqref{QI_def} and the interface charge invariant \eqref{QI_lambda}  for a randomly generated $N_c=Z=5$ model with a weakened link between sites $m=0$ and $m=1$ with six occupied bands. The components of the Hamiltonian are listed in the Supplemental Material [\onlinecite{supp_mat}]. Panel a) shows the band structure of the translationally invariant system (i.e.~for $\lambda = 1$) as a function of quasimomentum $k$ with the relevant gap marked by the black rectangle. Panel b) shows a zoom into this gap with dots depicting the energy of localized eigenstates of the finite Hamiltonian  with $N_Z=100$ unit cells as function of the link strength $\lambda$. The dashed line indicates the value of the chemical potential. Panel c) shows the lineshapes of the boundary charges on the left and right side of the impurity ($Q_B^{(L/R)}$), the  interface charge $Q_I^{\text{def}}$ and the invariant $Q_I$, all as functions of the link strength $\lambda$. Note that $Q_I$ and $Q_I^{\text{def}}$ perfectly overlap.}
    \label{fig:linkWeakeningExample}
\end{figure}

Fig.~\ref{fig:linkWeakeningExample} shows a comparison of the interface charge $Q_I^{\text{def}}$ (computed in terms of \eqref{QI_def} as described above) with the corresponding winding number invariant $Q_I$ given in Eq.~(\ref{QI_lambda}) for a randomly generated $N_c=Z=5$ model with six occupied bands (the Hamiltonian parameters are listed in the Supplemental Material [\onlinecite{supp_mat}]).

Panel a) of Fig.~(\ref{fig:linkWeakeningExample}) shows the band structure of the translationally invariant system (i.e.~for $\lambda=1$) as a function of quasimomentum $k$ and the relevant gap as denoted by the black rectangle. 

Panel b) shows a zoom onto the relevant gap with the chemical potential (dashed black line) and the energy of eigenstates of the finite system with $N_Z=100$ unit cells as a function of the link strength $\lambda$. One can see that in total, as the strength of the link is weakened, two eigenmodes leave the sixth band of the system turning into bound states.

Panel c) shows the boundary charges $Q_B^{(L)}$ and $Q_B^{(R)}$ on the left ($m \leq 0$) and right ($m \geq 1$) sides of the impurity, the interface charge $Q_I^{\text{def}}$ computed on the basis of \eqref{QI_def}, and the winding number invariant $Q_I$, all as functions of the link strength $\lambda$. One can see that $Q_I^{\text{def}}$ and $Q_I$ agree throughout the range of $\lambda$, confirming the validity of Eq.~(\ref{QI_lambda}) for this random model.

Again, we report that we tested numerous random models and always found agreement of \eqref{QI_def} with Eq.~(\ref{QI_lambda}).

\section{Summary}
\label{sec:summary}

For generic one-dimensional insulators with $Z$ sites per unit cell and $N_{c}$ channels (orbitals) per site we have developed a general theoretical framework for boundary and interface charge investigation, employing the method of boundary Green's functions. Using this approach, we represent the boundary and interface charges in terms of bulk Green's functions of the system, which proves particularly advantageous when dealing with multi-channel systems, in which the construction of exact eigenstates of the system with broken translational invariance is usually a rarely achievable goal due to the possibility of overlapping bands. 

We use this representation of the boundary charge in order to establish the topological invariant associated with the changes of boundary charge under lattice shifts. It was shown that upon a shift of the lattice as a whole by a single lattice site $Q_{B}$ (relative to the average density in the bulk $\bar\rho$) changes by an integer $-N_c\leq I\leq 0$, which can be expressed as a winding number of a particular combination of bulk Green's function components. This result is a natural generalization of our previous findings regarding single-channel systems, where $I$ was shown to take on values in $0$ or $-1$. The present finding proves the conjecture in Sec. VI A of Ref.~[\onlinecite{pletyukhov_etal_prb_20}]. The character of the bulk-boundary correspondence described by this invariant is not limited to any particular symmetry class and holds for arbitrary systems. This invariant quantifies the spectral flow of the boundary eigenvalue problem, the change by $-1$ in it indicating every time that an electron is taken away by an emerging edge state (which can happen at most $N_c$ times during one pumping cycle of the phase $\varphi$).

In addition to the quantization of $\Delta Q_{B}-\bar\rho$ we established the quantization of charge accumulated on local scattering centers, also known as interface charge. Specifically, we demonstrate that the charge accumulating either around  a single potential impurity or a weak link  is given by a corresponding winding number and is thus a topological invariant. This topological invariant also quantifies the spectral flow --- now of the interface eigenvalue problems. This observation generalizes the findings of Ref. [\onlinecite{Pletyukhov_etal3_2020}] (see Appendix C therein) and provides an analytical proof of the nearsightedness principle for generic non-interacting one-dimensional tight-binding models with translational invariance to which a single impurity is added.

In future work [\onlinecite{Pias_new}] it would be also interesting to extend the obtained results to a class of models with longer than nearest-neighbor range hoppings. Although longer ranged hoppings can always be rewritten effectively in terms of nearest-neighbor hopping by increasing the number of channels per site, it is of interest to avoid this artificial procedure and to find winding number expressions for the invariants including longer ranged hoppings.

\section{Acknowledgments}
\label{sec:acknowledgments}

We acknowledge fruitful discussions  with J. Klinovaja and D. Loss on the subject of the boundary charge. The work was supported by the Deutsche Forschungsgemeinschaft via RTG 1995.

\begin{appendix}

\section{Properties of boundary Green's functions and boundary charge}
\label{app:green_func}

\subsection{Boundary charge expression}
\label{app:bc}

Let us introduce the following short-hand notation for the Bloch Hamiltonian defined in Eq.~\eqref{bloch_hk}
\begin{align}
    h_k  &= \left( \begin{array}{cc} A & b_k \\ b_{k}^{\dagger} & v_Z \end{array}\right), \label{bloch_hk_block0} \\
    h_k - (\omega + i \eta) &= \left( \begin{array}{cc} \bar{A} & b_k \\ b_{k}^{\dagger} & \bar{v}_Z \end{array}\right),
    \label{bloch_hk_block}
\end{align}
where $\bar{A} = A -(\omega +i\eta)$, $\bar{v}_Z = v_Z - (\omega+i \eta)$, 
\begin{align}
b_k = - \left( \begin{array}{c} t_Z  e^{-ik} \\ 0 \\ \vdots \\ 0 \\ t_{Z-1}^{\dagger} \end{array} \right), \quad b_k^{\dagger} =  -\left(  e^{i k} t_Z^{\dagger}, 0,\dots,0, t_{Z-1} \right)
\label{bk_def}
\end{align}
and 
\begin{equation}
    A = 
    \begin{pmatrix}
        v_1 & -t_1^{\dagger} \\
        -t_1 & v_2 & \ddots \\
             & \ddots & \ddots  & -t_{Z-2}^{\dagger} \\
             &        & -t_{Z-2} & v_{Z-1}
    \end{pmatrix}.
\end{equation}

Using the matrix identity 
\begin{widetext}
\begin{align}
 \left( \begin{array}{cc} M_{11} & M_{12} \\ M_{21} & M_{22} \end{array}\right)^{-1} =  \left( \begin{array}{cc} (M_{11} - M_{12} M_{22}^{-1} M_{21})^{-1} &  -M_{11}^{-1} M_{12}  (M_{22} - M_{21} M_{11}^{-1} M_{12})^{-1} \\ -  (M_{22} - M_{21} M_{11}^{-1} M_{12})^{-1} M_{21} M_{11}^{-1}  & (M_{22} - M_{21} M_{11}^{-1} M_{12})^{-1} \end{array}\right),
 \label{matr_iden}
\end{align}
\end{widetext}
where $M_{11}$ and $M_{22}$ are square invertible matrices, which may eventually have different sizes, we evaluate the inverse of \eqref{bloch_hk_block} and find
\begin{align}
\langle Z | \frac{1}{\omega+ i \eta - h_k} |Z \rangle &=  - (\bar{v}_Z - b_{k}^{\dagger} \bar{A}^{-1} b_k)^{-1}, \\
 \langle Z | \frac{1}{\omega + i \eta - h_k} |j \rangle  &=  (\bar{v}_Z - b_{k}^{\dagger} \bar{A}^{-1} b_k)^{-1} (b_{k}^{\dagger} \bar{A}^{-1})_j , \\
  \langle j | \frac{1}{\omega + i \eta - h_k} |Z \rangle &= (\bar{A}^{-1}  b_k)_j   (\bar{v}_Z - b_{k}^{\dagger} \bar{A}^{-1} b_k)^{-1} ,  
\end{align}
where the last two relations are valid for $1 \leq j \leq Z-1 $. Using \eqref{bk_def}, we express
\begin{align}
 b_k^{\dagger} \bar{A}^{-1} b_k &=  t_Z^{\dagger} \bar{A}^{-1}_{1,1} t_Z + t_{Z-1} \bar{A}^{-1}_{Z-1,Z-1} t_{Z-1}^{\dagger} \nonumber \\
  & + e^{i k} t_Z^{\dagger} \bar{A}^{-1}_{1,Z-1} t_{Z-1}^{\dagger} +t_{Z-1} \bar{A}^{-1}_{Z-1,1} t_Z e^{-ik} , \\
 (b_{k}^{\dagger} \bar{A}^{-1})_j  &= -  e^{i k}  t_Z^{\dagger} \bar{A}^{-1}_{1,j} - t_{Z-1} \bar{A}^{-1}_{Z-1,j}, \\
 (\bar{A}^{-1} b_k)_j &= -\bar{A}^{-1}_{j,1} t_Z e^{-ik} - \bar{A}^{-1}_{j,Z-1} t_{Z-1}^{\dagger}.
\end{align}
Introducing the notations
\begin{align}
 c &= t_Z^{\dagger} \bar{A}^{-1}_{1,Z-1} t_{Z-1}^{\dagger}, \label{c_or}\\
 c^{\dagger} &= t_{Z-1} \bar{A}^{-1}_{Z-1,1} t_{Z}, \label{c_dagger} \\
r &=  \bar{v}_Z -   t_Z^{\dagger} \bar{A}^{-1}_{1,1} t_Z - t_{Z-1} \bar{A}^{-1}_{Z-1,Z-1} t_{Z-1}^{\dagger}, \label{r_or} \\
m (k) &= \left( c e^{i k} + c^{\dagger} e^{-i k} - r\right)^{-1}, \\
F_n &= \int_{-\pi}^{\pi} \frac{d k}{2 \pi} m (k)  e^{i k n} ,
\label{Fn_def}
\end{align}
we establish the following relations for $j=Z$
\begin{align}
\label{eq:def_F0}
G_{0,0}^{(0)} (\omega) &=  F_0, \\
G_{0,m}^{(0)} (\omega) &= F_{-n}, \quad G_{m,0}^{(0)} (\omega) = F_{n},
\end{align}
as well as for $j \neq Z$
\begin{align}
G_{0,m}^{(0)} (\omega) &=  F_{-n+1} t_Z^{\dagger} \bar{A}^{-1}_{1,j} + F_{-n} t_{Z-1} \bar{A}^{-1}_{Z-1,j} , \label{eq:G0m_0} \\
G_{m,0}^{(0)} (\omega) &=  \bar{A}^{-1}_{j,1} t_Z  F_{n-1} + \bar{A}^{-1}_{j,Z-1} t_{Z-1}^{\dagger} F_n  .
\end{align}
Note that defining the hermitian conjugation in \eqref{c_dagger} we assume  that it does \textit{not} conjugate the spectral parameter $\omega + i \eta$. We also remark that from \eqref{eq:G0m_0} it follows for $m=1$, i.e. for $n=1$ and $j=1$,
\begin{align}
    G_{0,1}^{(0)} (\omega) = (F_{0} t_Z^{\dagger} \bar{A}^{-1}_{1,1} t_Z + F_{-1} c^{\dagger}) t_Z^{-1}.
    \label{eq:G01_0}
\end{align}

Using this representation we perform the sum
\begin{align}
& \sum_{m=1}^{\infty} G_{0,m}^{(0)} (\omega) G_{m,0}^{(0)} (\omega) = \sum_{n=1}^{\infty} \sum_{j=1}^Z G_{0,m}^{(0)} (\omega) G_{m,0}^{(0)} (\omega) \nonumber \\
&=\sum_{n=1}^{\infty} (F_{-n+1} t_Z^{\dagger} \bar{A}^{-2}_{1,1} t_Z F_{n-1} +F_{-n} t_{Z-1} \bar{A}_{Z-1,Z-1}^{-2} t_{Z-1}^{\dagger} F_n) \nonumber \\
&+\sum_{n=1}^{\infty} (F_{-n+1} t_Z^{\dagger} \bar{A}^{-2}_{1,Z-1} t_{Z-1}^{\dagger} F_n + F_{-n} t_{Z-1} \bar{A}^{-2}_{Z-1,1} t_Z F_{n-1}) \nonumber \\
&+ \sum_{n=1}^{\infty} F_{-n} F_n .
\label{sum_m_1}
\end{align}

With help of the differential identity
\begin{align}
    \frac{\partial \bar{A}^{-1}}{\partial \omega} = - \bar{A}^{-1} \frac{\partial \bar{A}}{\partial \omega} \bar{A}^{-1} = \bar{A}^{-2}
\end{align}
we simplify Eq.~\eqref{sum_m_1} to
\begin{align}
& \sum_{m=1}^{\infty} G_{0,m}^{(0)} (\omega) G_{m,0}^{(0)} (\omega) \nonumber \\
=&\sum_{n=1}^{\infty} (F_{-n+1} \frac{\partial c}{\partial \omega} F_n + F_{-n}  \frac{\partial c^{\dagger}}{\partial \omega} F_{n-1} - F_{-n} \frac{\partial r}{\partial \omega}  F_n) \nonumber \\
&+ F_{0} t_Z^{\dagger} \frac{\partial \bar{A}^{-1}_{1,1}}{\partial \omega} t_Z F_{0}.
\label{sum_m_2}
\end{align}

Using the identities
\begin{align}
      F_{-n+1} c + F_{-n-1} c^{\dagger} -  F_{-n}  r  &= \delta_{n,0}
 \label{Fc_ident}, \\
  c F_{n+1} +c^{\dagger} F_{n-1}  -    r F_n &= \delta_{n,0},
 \label{cF_ident}
\end{align}
we find two equivalent representations
\begin{align}
& \sum_{m=1}^{\infty} G_{0,m}^{(0)} (\omega) G_{m,0}^{(0)} (\omega) \nonumber \\
=& - \sum_{n=1}^{\infty} ( \frac{\partial F_{-n+1}}{\partial \omega} c +   \frac{\partial F_{-n-1}}{\partial \omega} c^{\dagger} - \frac{\partial  F_{-n}}{\partial \omega} r )  F_n \nonumber \\
&+ F_{0} t_Z^{\dagger} \frac{\partial \bar{A}^{-1}_{1,1}}{\partial \omega} t_Z F_{0} + F_{-1}  \frac{\partial c^{\dagger}}{\partial \omega} F_{0}\nonumber \\
=&   F_{0} t_Z^{\dagger} \frac{\partial \bar{A}^{-1}_{1,1}}{\partial \omega} t_Z F_{0} +   \frac{\partial (F_{-1} c^{\dagger})}{\partial \omega} F_{0} -\frac{\partial F_{0}}{\partial \omega} c F_{1}
\label{sum_m_2a} 
\end{align}
and
\begin{align}
& \sum_{m=1}^{\infty} G_{0,m}^{(0)} (\omega) G_{m,0}^{(0)} (\omega) \nonumber \\
=& -\sum_{n=1}^{\infty} F_{-n} (c \frac{\partial F_{n+1}}{\partial \omega}  +  c^{\dagger} \frac{\partial  F_{n-1}}{\partial \omega} -  r \frac{\partial F_n}{\partial \omega}  ) \nonumber \\
&+ F_{0} t_Z^{\dagger} \frac{ \partial \bar{A}^{-1}_{1,1}}{\partial \omega} t_Z F_{0} + F_0 \frac{\partial c}{\partial \omega} F_1 \nonumber \\
=&  F_{0} t_Z^{\dagger} \frac{ \partial \bar{A}^{-1}_{1,1}}{\partial \omega} t_Z F_{0} + F_0 \frac{\partial (c F_1 )}{\partial \omega}  -F_{-1} c^{\dagger} \frac{\partial  F_{0}}{\partial \omega} .
\label{sum_m_3}
\end{align}
Introducing
\begin{align}
p =  \bar{v}_Z +   t_Z^{\dagger} \bar{A}^{-1}_{1,1} t_Z - t_{Z-1} \bar{A}^{-1}_{Z-1,Z-1} t_{Z-1}^{\dagger} = r + 2t_Z^{\dagger} \bar{A}^{-1}_{1,1} t_Z
\label{p_or}
\end{align}
we show that
\begin{align}
& \sum_{m=1}^{\infty} G_{0,m}^{(0)} (\omega) G_{m,0}^{(0)} (\omega) = \frac12 F_{0}  \frac{\partial p}{\partial \omega}  F_{0} - \frac12   \frac{\partial  F_0}{\partial \omega} \nonumber \\
&+   \frac12 \frac{\partial (F_{-1} c^{\dagger} - F_1 c)}{\partial \omega} F_{0} -\frac12 \frac{\partial F_{0}}{\partial \omega} (c F_1 -c^{\dagger} F_{-1}) ,
\end{align}
\begin{align}
& \sum_{m=1}^{\infty} G_{0,m}^{(0)} (\omega) G_{m,0}^{(0)} (\omega) = \frac12 F_{0}  \frac{\partial p}{\partial \omega}  F_{0} - \frac12  \frac{\partial  F_0}{\partial \omega} \nonumber \\
& + \frac12 F_0 \frac{\partial (c F_1 -c^{\dagger} F_{-1})}{\partial \omega}  -\frac12 (F_{-1} c^{\dagger}-F_1 c) \frac{\partial  F_{0}}{\partial \omega}.
\end{align}
Summing halves of each expression we obtain
\begin{align}
& \text{tr} \left\{ [G_{0,0}^{(0)} ]^{-1} \sum_{m=1}^{\infty} G_{0,m}^{(0)} (\omega) G_{m,0}^{(0)} (\omega) \right\}\nonumber \\
&= \frac12 \text{tr} \left\{  \frac{\partial p}{\partial \omega}  F_{0} \right\}- \frac12 \text{tr} \left\{ F_0^{-1 } \frac{\partial  F_0}{\partial \omega} \right\} \nonumber \\
&   -\frac14 \text{tr} \left\{ F_0^{-1 }\frac{\partial F_{0}}{\partial \omega} (c F_1 -c^{\dagger} F_{-1}) \right\}  \nonumber \\
&-\frac14  \text{tr} \left\{ (F_{-1} c^{\dagger}-F_1 c) \frac{\partial  F_{0}}{\partial \omega}  F_0^{-1 }\right\}.
\label{fried_spec}
\end{align}

\subsection{Identities relating components of bulk Green's functions}

In this subsection we review properties of the matrices $F_n$ defined in Eq.~(\ref{Fn_def}) and and prove some useful identities associated with them.

Rewriting the identity in Eq.~\eqref{cF_ident} in the matrix form we find
\begin{align}
    F^{-1}_{n,n'} F_{n',n''} = \delta_{n,n''},
\end{align}
with $F_{n',n''} \equiv F_{n'-n''}$ and
\begin{align}
F^{-1}_{n,n'} = c \delta_{n+1,n'} + c^{\dagger} \delta_{n-1,n'} -r \delta_{n,n'}.
\end{align}
We view $F^{-1}_{n,n'}$ as a Hamiltonian of the effective tight-binding model, $F_{n',n''}$ being the corresponding translationally invariant Green's function. Perturbing this model by cutting the link between the zeroth and first sites,
\begin{align}
    (F_S)^{-1}_{n,n'} &= F^{-1}_{n,n'} + (V_S)_{n,n'} \\
    &= F^{-1}_{n,n'} -c \delta_{n,0} \delta_{n',1}- c^{\dagger} \delta_{n,1} \delta_{n',0},
\end{align}
we split this system into two disconnected parts. Since the two parts are completely independent we find $(F_S)_{n,n'} =0$ for $n \leq 0$, $n' \geq 1$ and for $n \geq 1$, $n' \leq 0$. The function $F_S$ is called the surface Green's function and obeys the following Dyson equations
\begin{align}
    F_S &= F - F V_S F_S = F - F_S V_S F, \\
    (F_S)_{n,n'} &= F_{n,n'} + F_{n,0} c (F_S)_{1,n'}+ F_{n,1} c^{\dagger} (F_S)_{0,n'} \\
    &= F_{n,n'} +(F_S)_{n,0} c F_{1,n'}+ (F_S)_{n,1} c^{\dagger} F_{0,n'}.
\end{align}
Choosing $n=0$, $n'=1$ we find
\begin{align}
    0 = F_{-1} + F_{0} c (F_S)_{1,1} = F_{-1} + (F_S)_{0,0} c F_0 
    \label{idF1}
\end{align}
and choosing $n=1$, $n'=0$ we find
\begin{align}
    0 = F_1 + F_0 c^{\dagger} (F_S)_{0,0} = F_1 + (F_S)_{1,1} c^{\dagger} F_{0}. \label{idF2}
\end{align}
Comparing the two expressions, we find the following identities 
\begin{align}
    F_0 c^{\dagger} F_{-1} &=   F_1 c F_0 , \label{sgfL}  \\
    F_0 c F_1 &= F_{-1} c^{\dagger} F_0 . \label{sgfR}
\end{align}

Next, focusing on the two semi-infinite Green's functions $(F_R)_{n,n'} = (F_S)_{n \geq 1,n' \geq 1}$, $(F_L)_{n,n'} = (F_S)_{n\leq 0,n'\leq 0} $,
and using the condition that they remain invariant under adding one more site to the corresponding semi-infinite models, i.e.
\begin{align}
    F_R = \left( \begin{array}{cc} F_R^{-1} & c^{\dagger} \\ c & -r \end{array}\right)^{-1} , \quad F_L = \left( \begin{array}{cc} -r & c^{\dagger} \\ c & F_L^{-1} \end{array}\right)^{-1} ,
\end{align}
we compare the $(1,1)$ matrix elements on both sides of the first relation (in the lower right corner), and the $(0,0)$ matrix elements on both sides of the second relation (in the upper left corner). To this end we use Eq.~\eqref{matr_iden}. We find
\begin{align}
    (F_S)_{1,1} &= (-r-c (F_S)_{1,1} c^{\dagger})^{-1}, \\
    (F_S)_{0,0} &= (- r - c^{\dagger} (F_S)_{0,0} c)^{-1}.
\end{align}
Eliminating $(F_S)_{1,1}$, $(F_S)_{0,0}$, and $r$ with the help of Eqs. \eqref{idF1}, \eqref{idF2}, \eqref{cF_ident} we obtain the following quadratic matrix equations
\begin{align}
    (c^{\dagger} F_{-1})^2 - c^{\dagger} F_{-1} - c^{\dagger} F_0 c F_0 &=0, \label{cdF_eq_sq} \\
    (F_1 c)^2  -  F_{1} c  - F_0 c^{\dagger} F_0 c  &=0, \label{Fc_eq_sq} \\
    (c F_1)^2 - c F_1 - c F_0 c^{\dagger} F_0 &=0, \\
    (F_{-1} c^{\dagger})^2 - F_{-1} c^{\dagger} - F_0 c F_0 c^{\dagger}  &=0.
\end{align}
Their solutions yields
\begin{align}
    c^{\dagger} F_{-1} &= \frac{1-\sqrt{1+ 4c^{\dagger} F_0 c F_0}}{2}, \\
    F_1 c &= \frac{1-\sqrt{1+ 4 F_0 c^{\dagger} F_0 c  }}{2}, \\
    c F_1 &=\frac{1-\sqrt{1+ 4 c F_0 c^{\dagger} F_0  }}{2}, \\
    F_{-1} c^{\dagger} &= \frac{1-\sqrt{1+4  F_0 c F_0 c^{\dagger} }}{2},
\end{align}
where the sign choice for the square roots is substantiated by the perturbative expansion of $F_1$ and $F_{-1}$ in $c$, $c^{\dagger}$. 

Combining Eq.~\eqref{Fc_eq_sq} with Eq.~\eqref{sgfL}, we show that
\begin{align}
    F_0 = (1-F_{1} c) (F_0 - F_1 F_0^{-1} F_{-1}) .
    \label{idF0_soph}
\end{align}

Analogous identities hold for the shifted system, that is for the tilded functions.

\subsection{Derivation of \eqref{wn_I}}
\label{app:shift_QB}

Performing a shift of the lattice leftwards by one site, as suggested in Sec.~\ref{sec:shift_model}, we effectively redefine the unit cell (i.e. making it begin from site $j=2$ and terminating it after site $j=1$). The corresponding Bloch Hamiltonian defined in Eq.~\eqref{bloch_hk_block} changes to 
\begin{align}
    \tilde{h}_k - (\omega + i \eta) = \left( \begin{array}{cc} \tilde{A} & \tilde{b}_k \\ \tilde{b}_{k}^{\dagger} & \bar{v}_1 \end{array}\right),
    \label{tilde_bloch_hk_block}
\end{align}
where $\bar{v}_1 = v_1 - (\omega+i \eta)$ and
\begin{align}
\tilde{b}_k = - \left( \begin{array}{c} t_1  e^{-ik} \\ 0 \\\vdots \\0\\ t_{Z}^{\dagger} \end{array} \right), \quad \tilde{b}_k^{\dagger} =  -\left(  e^{i k} t_1^{\dagger}, 0,\ldots,0, t_{Z} \right).
\label{tilde_bk_def}
\end{align}
The matrix $\tilde{A}$ has the structure similar to that of $\bar{A}$ --- they differ from each other only by the labelling of sites. Moreover, they share a common block matrix $a$ of the size $N_c (Z-2) \times N_c (Z-2)$. This becomes obvious in the following representation
\begin{align}
    \bar{A} = \left( \begin{array}{cc} \bar{v}_1 & -T_1^{\dagger} \\ -T_1 & a \end{array}\right) , \quad  \tilde{A} = \left( \begin{array}{cc} a & -T_{Z-1} \\ -T_{Z-1}^{\dagger} & \bar{v}_Z \end{array} \right),
\end{align}
with
\begin{align}
T_1 &=  \left( \begin{array}{c} t_1   \\ \vdots \\ 0 \end{array} \right), \quad T_1^{\dagger} =  \left(   t_1^{\dagger}, \ldots, 0 \right),
\label{T1} \\
T_{Z-1} &=  \left( \begin{array}{c} 0  \\ \vdots \\ t_{Z-1}^{\dagger} \end{array} \right), \quad T_{Z-1}^{\dagger} =  \left(  0, \ldots, t_{Z-1} \right).
\label{TZM1}
\end{align}

Using the identity in Eq.~\eqref{matr_iden} and the additional relation
\begin{align}
    &-M_{11}^{-1} M_{12} (M_{22} - M_{21} M_{11}^{-1} M_{12})^{-1} 
    \nonumber \\
    =& -  ( M_{11}- M_{12} M_{22}^{-1} M_{21}  )^{-1} M_{12} M_{22}^{-1},
\end{align}
we evaluate
\begin{align}
    \bar{A}_{1,1}^{-1} &= (\bar{v}_1 - t_1^{\dagger} a_{2,2}^{-1} t_1)^{-1} \equiv V_1^{-1}, \\
    \bar{A}_{Z-1,Z-1}^{-1} &= a_{Z-1,Z-1}^{-1} +a_{Z-1,2}^{-1} t_1 V_1^{-1} t_1^{\dagger} a_{2,Z-1}^{-1}, \\
    \bar{A}_{1,Z-1}^{-1} &= V_1^{-1} t_1^{\dagger} a_{2,Z-1}^{-1}, \\
    \bar{A}_{Z-1,1}^{-1} &= a_{Z-1,2}^{-1} t_1 V_1^{-1}, 
\end{align}
and 
\begin{align}
    \tilde{A}_{1,1}^{-1} &=a_{2,2}^{-1} + a_{2,Z-1}^{-1} t_{Z-1}^{\dagger} V_Z^{-1} t_{Z-1} a^{-1}_{Z-1,2}, \\
    \tilde{A}_{Z-1,Z-1}^{-1} &= (\bar{v}_Z - t_{Z-1} a_{Z-1,Z-1}^{-1} t_{Z-1}^{\dagger})^{-1} \equiv V_Z^{-1} ,\\
    \tilde{A}_{1,Z-1}^{-1} &= a_{2,Z-1}^{-1} t_{Z-1}^{\dagger} V_Z^{-1}, \\
    \tilde{A}_{Z-1,1}^{-1} &=  V_Z^{-1} t_{Z-1} a_{Z-1,2}^{-1}.
\end{align}
Note that we label the blocks of the matrix $a$ beginning with $2$ and ending with $Z-1$.

Using these representations we express Eqs.~\eqref{c_or}, \eqref{c_dagger}, \eqref{r_or}, and \eqref{p_or} as follows
\begin{align}
 c &= t_Z^{\dagger} V_1^{-1} \bar{t}_1^{\dagger} , \label{c_or_re} \\
 c^{\dagger} &=  \bar{t}_1 V_1^{-1}t_{Z} ,  \label{c_dagger_re} \\
r &=  V_Z -   t_Z^{\dagger} V^{-1}_{1} t_Z  -  \bar{t}_1 V_1^{-1} \bar{t}_1^{\dagger} , \label{r_or_re} \\
p &=  V_Z +   t_Z^{\dagger} V^{-1}_{1} t_Z - \bar{t}_1 V_1^{-1} \bar{t}_1^{\dagger} ,
\label{p_or_re}
\end{align}
where we have introduced the notation $\bar{t}_1 =  t_{Z-1} a_{Z-1,2}^{-1} t_1$.
In addition, we find analogs of these quantities for the shifted system
\begin{align}
 \tilde{c} &= t_1^{\dagger} \tilde{A}^{-1}_{1,Z-1} t_{Z}^{\dagger} = \bar{t}_1^{\dagger}  V_Z^{-1} t_{Z}^{\dagger} , \label{c_shift}\\
 \tilde{c}^{\dagger} &= t_{Z} \tilde{A}^{-1}_{Z-1,1} t_{1} = t_{Z} V_Z^{-1} \bar{t}_{1} , \label{c_dagger_shift} \\
\tilde{r} &=  \bar{v}_1 -   t_1^{\dagger} \tilde{A}^{-1}_{1,1} t_1 - t_{Z} \tilde{A}^{-1}_{Z-1,Z-1} t_{Z}^{\dagger} \nonumber \\
&= V_1- \bar{t}_1^{\dagger}  V_Z^{-1}\bar{t}_1 - t_{Z} V^{-1}_{Z} t_{Z}^{\dagger} , \label{r_shift} \\
\tilde{p} &=  \bar{v}_1 +   t_1^{\dagger} \tilde{A}^{-1}_{1,1} t_1 - t_{Z} \tilde{A}^{-1}_{Z-1,Z-1} t_{Z}^{\dagger} \nonumber \\
&= 2 \bar{v}_1 - (V_1  -  \bar{t}_1^{\dagger}   V_Z^{-1} \bar{t}_1 + t_{Z} V^{-1}_{Z} t_{Z}^{\dagger}) \equiv 2 \bar{v}_1 -\tilde{p}_* .
\label{p_shift}
\end{align}
As one can see, they all are expressed just in terms of the four matrices $V_1$, $V_Z$, $\bar{t}_1$, and $t_Z$ (besides $\tilde{p}$, which has an extra contribution $2 \bar{v}_1$).

Defining
\begin{align}
\tilde{m} (k) &= \left( \tilde{c} e^{i k} + \tilde{c}^{\dagger} e^{-i k} - \tilde{r}\right)^{-1}, \\
\tilde{F}_n &= \int_{-\pi}^{\pi} \frac{d k}{2 \pi} \tilde{m} (k) e^{i k n},
\label{tilde_Fn_def}
\end{align}
and using Eqs.~\eqref{eq:QFE} and \eqref{fried_spec} for both the initial and the shifted systems we evaluate 
\begin{align}
    \Delta Q'_B &= \tilde{Q}'_B - Q'_B \nonumber \\
    &= \frac{1}{\pi} \int d \omega \,\, \Theta (\mu-\omega)\, \text{Im} \, B \label{K_contr} \\
    &- \frac{1}{\pi}  \int d \omega \,\, \Theta (\mu-\omega) \, \text{Im  tr} \, \tilde{F}_0 , \label{rho1_counter}
\end{align}
where
\begin{align}
    B=& - \frac12 \text{tr} \left\{  \tilde{F}_0^{-1 } \frac{\partial  \tilde{F}_0}{\partial \omega} \right\} +\frac12 \text{tr} \left\{  \tilde{F}_{0} \frac{\partial }{\partial \omega} \left( \tilde{\Lambda}^{\dagger} \tilde{F}_0^{-1 } + \tilde{F}_0^{-1 } \tilde{\Lambda}\right)\right\}   \nonumber \\
&+ \frac12 \text{tr} \left\{  F_0^{-1 } \frac{\partial  F_0}{\partial \omega} \right\} + \frac12 \text{tr} \left\{ F_0 \frac{\partial}{\partial \omega} \left( \Lambda  F_0^{-1 } + F_0^{-1 } \Lambda^{\dagger} \right) \right\}  ,
\end{align}
and
\begin{align}
\tilde{\Lambda} &= \frac{\tilde{F}_{-1} \tilde{c}^{\dagger}-\tilde{F}_{1} \tilde{c} }{2} -\tilde{F}_{0} \frac{\tilde{p}_*}{2} , \\
\label{eq:def_Lambda}
\Lambda &= \frac{c^{\dagger} F_{-1}-c F_{1}}{2} -\frac{p}{2} F_{0} .
\end{align}

Noticing that the contribution in Eq.~\eqref{rho1_counter} equals $\rho_1^{(0)}$ (since $\tilde{F}_0 = \tilde{G}_{0,0}^{(0)} =G_{1,1}^{(0)}$) we achieve exact cancellation of the corresponding term in \eqref{inv_deltaQ}. Thereby we get
\begin{align}
    I = \frac{1}{\pi} \int d \omega \,\, \Theta (\mu-\omega) \, \text{Im}\, B .
\end{align}

On the basis of Eqs.~\eqref{sgfL}, \eqref{sgfR} we conclude that
\begin{align}
    \tilde{F}_0 \tilde{\Lambda}^{\dagger} &=\tilde{\Lambda} \tilde{F}_0 , \label{LdLd_exch} \\
    F_0 \Lambda &= \Lambda^{\dagger} F_0. \label{LL_exch}
\end{align}
This observation allows us to write down $B$ in a more compact form
\begin{align}
    B=& - \frac12 \text{tr} \left\{  \tilde{F}_0^{-1 } \frac{\partial  \tilde{F}_0}{\partial \omega} \right\} + \text{tr} \left\{  \tilde{F}_{0} \frac{\partial }{\partial \omega} \left(\tilde{F}_0^{-1 } \tilde{\Lambda}\right)\right\}   \nonumber \\
&+ \frac12 \text{tr} \left\{  F_0^{-1 } \frac{\partial  F_0}{\partial \omega} \right\} +  \text{tr} \left\{ F_0 \frac{\partial}{\partial \omega} \left( \Lambda  F_0^{-1 }  \right) \right\}  .
\end{align}
With help of the identities (see the next section for the proof)
\begin{align}
\Lambda  &=  t_Z^{\dagger} \tilde{\Lambda} (t_Z^{\dagger})^{-1}, \label{idL2}\\
t_Z^{\dagger} \tilde{F}_0  t_Z &= \left(  \Lambda^2 - \frac14 \right) F_0^{-1},
\label{idL3}
\end{align}
we express
\begin{align}
    B &=  \text{tr} \left\{  F_0^{-1 } \frac{\partial  F_0}{\partial \omega} \right\} - \text{tr} \left\{  (\Lambda-\frac12)^{-1}   \frac{\partial }{\partial \omega}(\Lambda - \frac12) \right\}  \\
    &= \frac{\partial}{\partial \omega} \ln \det \left[ F_0 (\Lambda - \frac12)^{-1} \right] \\
    &= -\frac{\partial}{\partial \omega} \ln \det \left[ F_0^{-1} (\Lambda^{\dagger} - \frac12) \right],
\end{align}
where we have applied Jacobi's formula in the second step. Taking into account that $F_0 = G_{0,0}^{(0)}$ as well as the relation
\begin{align}
    G_{0,1}^{(0)} =  (\frac12 -\Lambda^{\dagger} ) t_Z^{-1},
    \label{G01expr}
\end{align}
following from Eq.~\eqref{eq:G01_0}, we finally obtain Eq.~\eqref{wn_I}.

\subsection{Identities relating components of boundary Green's functions}

Our goal is to prove the relations in Eqs.~\eqref{idL2}, \eqref{idL3}.

Let us treat the shifted system as the union of the first $Z-1$ sites and the rest of the semi-infinite lattice (which coincides with the initial system up to relabeling of sites $m \to m-Z+1$). They are coupled with each other by the hoppings $-t_Z$ and $-t_Z^{\dagger}$, which occur between sites $Z-1$ and $Z$. Formally, this separation is expressed as
\begin{align}
    \tilde{G}_{m,m'}^{-1}  &= (\tilde{G}')_{m,m'}^{-1} \nonumber \\
    &+ t_Z \delta_{m,Z} \delta_{m',Z-1}+  t_Z^{\dagger} \delta_{m,Z-1} \delta_{m',Z} , \label{sep_shift} \\
    (\tilde{G}')_{m,m'}^{-1} &= -\tilde{A}_{m,m'} \delta_{ 1 \leq m,m' \leq Z-1} \nonumber \\
    &+ G_{m-Z+1,m'-Z+1}^{-1} \delta_{m,m' \geq Z},
\end{align}
where $\tilde{G}'$ is the Green's function of the system in the absence of coupling between its two subsystems. It is convenient to rewrite Eq.~\eqref{sep_shift} in the form of the Dyson equation
\begin{align}
    (\tilde{G}')_{m,m'} = \tilde{G}_{m,m'} &+ \tilde{G}_{m,Z} t_Z (\tilde{G}')_{Z-1,m'} \nonumber \\
    &+ \tilde{G}_{m,Z-1} t_Z^{\dagger}  (\tilde{G}')_{Z,m'}.
\end{align}

Choosing $m=Z$ and $m' = Z-1,Z$, and using that $(\tilde{G}')_{Z,Z-1} =(\tilde{G}')_{Z-1,Z}=0$ (the case of two isolated subsystems), we obtain the following identities
\begin{align}
    0 &= \tilde{G}_{Z,Z-1} - \tilde{G}_{Z,Z} t_Z V^{-1}_{Z} , \label{Zm1_Z_equal} \\
    G_{1,1} &= \tilde{G}_{Z,Z} + \tilde{G}_{Z,Z-1} t_Z^{\dagger} G_{1,1}. \label{ZZ_equal} 
\end{align}

Using the general expression Eq.~\eqref{BGF_main} for the boundary Green's functions we express
\begin{align}
    \tilde{G}_{Z,Z} &= \tilde{G}_{Z,Z}^{(0)} -  \tilde{G}_{Z,0}^{(0)} [\tilde{G}_{0,0}^{(0)}]^{-1} \tilde{G}_{0,Z}^{(0)} \\
    &= \tilde{F}_0 - \tilde{F}_1 \tilde{F}_0^{-1} \tilde{F}_{-1}.
    \label{tGZZ_v2}
\end{align}
By virtue of the tilded analog of the identity in Eq.~\eqref{idF0_soph} we establish
\begin{align}
    \tilde{G}_{Z,Z}^{-1} &= \tilde{F}_0^{-1} (1- \tilde{F}_{1} \tilde{c}). 
    \label{tGZZ1} 
\end{align}
To prove Eq.~\eqref{idL2} we represent $\tilde{G}_{Z,Z-1}$ by means of the identity in Eq.~\eqref{BGF_main}
\begin{align}
    \tilde{G}_{Z,Z-1} &= \tilde{G}_{Z,Z-1}^{(0)} -  \tilde{G}_{Z,0}^{(0)} [\tilde{G}_{0,0}^{(0)}]^{-1} \tilde{G}_{0,Z-1}^{(0)} \nonumber \\
    &= G_{1,0}^{(0)} - \tilde{F}_1 \tilde{F}_0^{-1} \tilde{G}_{0,Z-1}^{(0)},
    \label{GZZm1_v2}
\end{align}
where we have used $\tilde{G}_{m-1,m'-1}^{(0)}=G_{m,m'}^{(0)}$, and hence $\tilde{G}_{Z,Z-1}^{(0)} = G_{Z+1,Z}^{(0)} = G_{1,0}^{(0)} $. In turn, representing $\tilde{G}_{0,Z-1}^{(0)}$ with help of the tilded analog of Eq.~\eqref{eq:G0m_0}, we obtain
\begin{align}
\tilde{G}_{0,Z-1}^{(0)} &=  \tilde{F}_{0} \tilde{t}_Z^{\dagger} \tilde{A}^{-1}_{1,Z-1} + \tilde{F}_{-1} \tilde{t}_{Z-1} \tilde{A}^{-1}_{Z-1,Z-1}  \\
&= (\tilde{F}_{0} \tilde{c}+ \tilde{F}_{-1}  t_Z V_Z^{-1} t_Z^{\dagger}) 
(t_Z^{\dagger})^{-1}, \label{G0Zm1_v1}
\end{align}
where we used have $\tilde{t}_Z = t_1$ and $\tilde{t}_{Z-1} = t_Z$. Combining Eqs.~\eqref{GZZm1_v2}, \eqref{G0Zm1_v1} with Eqs.~\eqref{Zm1_Z_equal}, \eqref{tGZZ_v2}, and \eqref{G01expr} we obtain the relation
\begin{align}
      (t_Z^{\dagger})^{-1} (\frac12 -\Lambda ) t_Z^{\dagger} &= \tilde{F}_{0} t_Z V^{-1}_{Z} t_Z^{\dagger}+ \tilde{F}_1 \tilde{c} \\
      &=  \frac{1 }{2} - \tilde{\Lambda} ,
\end{align}
which proves Eq.~\eqref{idL2}.
To prove Eq.~\eqref{idL3} we eliminate $\tilde{G}_{Z,Z-1}$ from Eqs.~\eqref{Zm1_Z_equal} and \eqref{ZZ_equal}. We obtain
\begin{align}
    G_{1,1} = \left(  \tilde{G}_{Z,Z}^{-1}- t_Z V^{-1}_{Z} t_Z^{\dagger} \right)^{-1}.
    \label{eq1:G11}
\end{align}
Then we substitute Eq.~\eqref{tGZZ1} into this relation and obtain
\begin{align}
    G_{1,1} &= \left(    1- \tilde{F}_{1} \tilde{c} -\tilde{F}_0  t_Z V^{-1}_{Z} t_Z^{\dagger} \right)^{-1} \tilde{F}_0 \\
    &= \left(    \frac12 + \tilde{\Lambda} \right)^{-1} \tilde{F}_0 =  (t_Z^{\dagger})^{-1} \left( \frac12 +  \Lambda  \right)^{-1} t_Z^{\dagger} \tilde{F}_0 .
    \label{eq2:G11}
\end{align}
On the other hand, on the basis of Eq.~\eqref{BGF_main}, it holds
\begin{align}
    G_{1,1} &= G_{1,1}^{(0)} - G_{1,0}^{(0)} [G_{0,0}^{(0)}]^{-1} G_{0,1}^{(0)} \\
    &= \tilde{F}_0 - (t_Z^{\dagger})^{-1} (\Lambda - \frac12) F_0^{-1} (\Lambda^{\dagger} - \frac12) t_Z^{-1} \\
    &= (t_Z^{\dagger})^{-1} \left[ t_Z^{\dagger} \tilde{F}_0 t_Z -  (\Lambda - \frac12)^2  F_0^{-1} \right]t_Z^{-1} . \label{eq3:G11}
\end{align}
Comparing Eq.~\eqref{eq2:G11} with Eq.~\eqref{eq3:G11} we deduce Eq.~\eqref{idL3}.

\section{Green's functions of the interface model with a weakened link}
\label{app:cut_link}

Below we establish the Green's function of the model introduced in Sec. \ref{sec:weak_link}.

Eliminating $G_{0,m'} $ from Eq.~\eqref{g0m}
\begin{align}
G_{0,m'} &= [1- G_{0,1}^{(0)} t_{\lambda}]^{-1} [G_{0,m'}^{(0)}  + G_{0,0}^{(0)} t_{\lambda}^{\dagger} G_{1,m'} ], \label{presol_g0m}
\end{align}
and inserting this result into Eq.~(\ref{g1m}), we obtain the following linear equation for $G_{1,m'}$:
\begin{align}
G_{1,m'} &= G_{1,m'}^{(0)} +  G_{1,1}^{(0)} t_{\lambda}  [1- G_{0,1}^{(0)} t_{\lambda}]^{-1} G_{0,m'}^{(0)} \nonumber \\
&+  \{ G_{1,1}^{(0)} t_{\lambda} [1- G_{0,1}^{(0)} t_{\lambda}^{\dagger}]^{-1} G_{0,0}^{(0)} t_{\lambda}^{\dagger}  +   G_{1,0}^{(0)} t_{\lambda}^{\dagger} \} G_{1,m'} .
 \label{presol_g1m}
\end{align}
Its solution reads
\begin{align}
 G_{1,m'} &=  S^{-1} \{ G_{1,m'}^{(0)} +  G_{1,1}^{(0)} t_{\lambda}  [1- G_{0,1}^{(0)} t_{\lambda} ]^{-1} G_{0,m'}^{(0)} \} ,
 \label{sol_g1m}
 \end{align}
 where
 \begin{align}
 S &=  1 -    G_{1,0}^{(0)} t_{\lambda}^{\dagger} -G_{1,1}^{(0)} t_{\lambda} [1- G_{0,1}^{(0)} t_{\lambda} ]^{-1} G_{0,0}^{(0)} t_{\lambda}^{\dagger} .
 \label{S_eq}
\end{align}
Analogously we find
\begin{align}
    G_{0,m'} =& [1- G_{0,1}^{(0)} t_{\lambda}]^{-1}G_{0,0}^{(0)} t_{\lambda}^{\dagger} S^{-1} \nonumber \\
& \times \{  [ 1  -    G_{1,0}^{(0)} t_{\lambda}^{\dagger} ] (t_{\lambda}^{\dagger})^{-1} G_{0,0}^{(0) -1}   G_{0,m'}^{(0)} +   G_{1,m'}^{(0)} \} .
\label{sol_g0m}
\end{align}
We note the useful matrix identity
\begin{align}
& 1+ G_{0,0}^{(0)} t_{\lambda}^{\dagger} S^{-1} G_{1,1}^{(0)} t_{\lambda} [1- G_{0,1}^{(0)} t_{\lambda}]^{-1} \nonumber \\
& =  G_{0,0}^{(0)} t_{\lambda}^{\dagger} S^{-1} [ 1 -    G_{1,0}^{(0)} t_{\lambda}^{\dagger}]  (t_{\lambda}^{\dagger})^{-1} G_{0,0}^{(0) -1}   ,
\end{align}
which helps us achieve various representations for the above solutions.
We also recall that $G_{0,0}^{(0)} = F_0$, $G_{1,1}^{(0)} = \tilde{F}_0$, and $G_{0,1}^{(0)}$ can be expressed via Eq.~\eqref{G01expr}.

Using the identities in Eqs.~\eqref{LL_exch}, \eqref{idL3} we simplify the expression in Eq.~\eqref{S_eq} as follows
\begin{align}
S =(t_Z^{\dagger})^{-1}  \frac{  1  - (1-\lambda^2 )  \left( \frac12 - \Lambda \right)}{ 1 - (1-\lambda) (\frac12 -  \Lambda) } t_Z^{\dagger}  =  \frac{  1  - (1-\lambda^2 )  \left( \frac12 - \tilde{\Lambda} \right)}{ 1 - (1-\lambda) (\frac12 -  \tilde{\Lambda}) }.
\end{align}
Consequently, we find
\begin{align}
G_{1,m'} &=   \frac{ 1 - (1-\lambda) (\frac12 -  \tilde{\Lambda}) }{  1  - (1-\lambda^2 )  \left( \frac12 - \tilde{\Lambda} \right)}  G_{1,m'}^{(0)}
\nonumber \\
&+    \frac{1-\lambda}{  1 - (1-\lambda^2) (\frac12 - \tilde{\Lambda} )}  \tilde{F}_{0} t_{Z} G_{0,m'}^{(0)}, \\
G_{0,m'} =&  \frac{  1  -   (1-\lambda) (\frac12 - \Lambda^{\dagger})  }{1- (1-\lambda^2) (\frac12 - \Lambda^{\dagger})  }      G_{0,m'}^{(0)} \nonumber \\
&+  \frac{1- \lambda}{1- (1-\lambda^2) (\frac12 - \Lambda^{\dagger})  }   F_0 t_{Z}^{\dagger} G_{1,m'}^{(0)} .
\label{sol_g0m_2}
\end{align}

Next, we are interested in
\begin{align}
& (1-\lambda)^{-1} \sum_{m=-\infty}^{\infty} \text{tr} \{ G_{m,m} - G_{m,m}^{(0)} \} \nonumber  \\
&=  \sum_{m=-\infty}^{\infty} \text{tr} \{ G_{m,1}^{(0)} t_Z G_{0,m} +  G_{m,0}^{(0)} t_Z^{\dagger} G_{1,m} \} \nonumber \\
&= \sum_{m=-\infty}^{\infty} \text{tr} \{ G_{m,1}^{(0)} t_Z  \frac{  1  -   (1-\lambda) (\frac12 - \Lambda^{\dagger})  }{1- (1-\lambda^2) (\frac12 - \Lambda^{\dagger})  }      G_{0,m}^{(0)}  \} \nonumber \\
&+  \sum_{m=-\infty}^{\infty} \text{tr} \{ G_{m,1}^{(0)} t_Z \frac{1- \lambda}{1- (1-\lambda^2) (\frac12 - \Lambda^{\dagger})  }   F_0 t_{Z}^{\dagger} G_{1,m}^{(0)}  \} \nonumber \\
&+  \sum_{m=-\infty}^{\infty} \text{tr} \{   G_{m,0}^{(0)} t_Z^{\dagger}  \frac{ 1 - (1-\lambda) (\frac12 -  \tilde{\Lambda}) }{  1  - (1-\lambda^2 )  \left( \frac12 - \tilde{\Lambda} \right)}  G_{1,m}^{(0)} \} \nonumber \\
&+  \sum_{m=-\infty}^{\infty} \text{tr} \{   G_{m,0}^{(0)} t_Z^{\dagger}     \frac{1-\lambda}{  1 - (1-\lambda^2) (\frac12 - \tilde{\Lambda} )}  \tilde{F}_{0} t_{Z} G_{0,m}^{(0)}  \} .
\end{align}

Considering
\begin{align}
\sum_{m=-\infty}^{\infty} G_{0,m}^{(0)} G_{m,0}^{(0)} &= \int_{-\pi}^{\pi} \frac{d k}{2 \pi} \sum_{j=1}^Z  \langle Z | \frac{1}{\omega + i\eta - h_k} | j \rangle \nonumber \\
 \times \langle j |\frac{1}{ \omega + i \eta -h_{k}} | Z \rangle &= - \frac{\partial F_0}{\partial \omega}, \label{Gm_sum} \\
\sum_{m=-\infty}^{\infty} G_{1,m}^{(0)} G_{m,1}^{(0)}  &= \sum_{m=-\infty}^{\infty} \tilde{G}_{0,m-1}^{(0)} \tilde{G}_{m-1,0}^{(0)}=  - \frac{\partial \tilde{F}_0}{\partial \omega} ,
\\
\sum_{m=-\infty}^{\infty} G_{0,m}^{(0)} G_{m,1}^{(0)} &=  \int_{-\pi}^{\pi} \frac{d k}{2 \pi} e^{-i k} \sum_{j=1}^Z  \langle Z | \frac{1}{\omega+ i \eta - h_k} | j \rangle \nonumber \\
 \times \langle j |\frac{1}{ \omega + i \eta -h_{k}} | 1 \rangle &= - \frac{\partial G_{0,1}^{(0)}}{\partial \omega} =  \frac{\partial \Lambda^{\dagger}}{\partial \omega} t_Z^{-1} ,
\end{align}
we obtain
\begin{align}
& \sum_{m=-\infty}^{\infty} \text{tr} \{ G_{m,m} - G_{m,m}^{(0)} \} \nonumber \\
&= (1-\lambda)  \text{tr} \left\{ \frac{  1  -   (1-\lambda) (\frac12 - \Lambda^{\dagger})  }{1- (1-\lambda^2) (\frac12 - \Lambda^{\dagger})  }       \frac{\partial \Lambda^{\dagger}}{\partial \omega}  \right\} \nonumber \\
&- (1-\lambda)  \text{tr} \left\{  t_Z \frac{1- \lambda}{1- (1-\lambda^2) (\frac12 - \Lambda^{\dagger})  }   F_0 t_{Z}^{\dagger}  \frac{\partial \tilde{F}_0 }{\partial \omega}   \right\} \nonumber \\
&+ (1-\lambda) \text{tr} \left\{    \frac{ 1 - (1-\lambda) (\frac12 -  \Lambda) }{  1  - (1-\lambda^2 )  \left( \frac12 - \Lambda \right)}   \frac{\partial \Lambda}{\partial \omega} \right\} \nonumber \\
&- (1-\lambda)  \text{tr} \left\{  \frac{1-\lambda}{  1 - (1-\lambda^2) (\frac12 - \Lambda )}  t_Z^{\dagger} \tilde{F}_{0} t_{Z} \frac{\partial F_0}{\partial \omega} \right\}.
\end{align}

Using again the identities in Eqs.~\eqref{LL_exch}, \eqref{idL3} we derive the expression
\begin{align}
& \sum_{m=-\infty}^{\infty} \text{tr} \{ G_{m,m} - G_{m,m}^{(0)} \} \nonumber \\
&=     \frac12 \text{tr} \left\{    \frac{ 1 }{  \frac{1+\lambda^2 }{2 (1-\lambda^2 )}   +  \Lambda }   \frac{\partial \Lambda}{\partial \omega} \right\} +    \frac12 \text{tr} \left\{    \frac{ 1 }{  \frac{1+\lambda^2 }{2 (1-\lambda^2 )}   +  \Lambda^{\dagger} }   \frac{\partial \Lambda^{\dagger} }{\partial \omega} \right\} \nonumber \\
&=\frac{\partial}{\partial \omega}  \ln \det \left[ (1- \lambda^2) \Lambda + \frac{1+ \lambda^2}{2} \right] ,
\end{align}
from which follows Eq.~\eqref{QI_lambda}.

\section{Topological invariants in the single-channel case}
\label{app:single_ch}

In this section we show how to recover the earlier obtained results for the single-channel case from the presently analysed multichannel expressions.

\subsection{Boundary charge invariant}
\label{app:single_ch_boundary}

In the single-channel case, $t_j$ and $v_j$ are scalar parameters. As explained in Ref.~[\onlinecite{pletyukhov_etal_prb_20}], by an appropriate gauge choice, one can make all hoppings real-valued, i.e. $t_j = t_j^*$. Therefore, $c= c^* = \frac{1}{\det \bar{A}} \prod_{j=1}^Z t_j $, $F_1= F_{-1}$, and
\begin{align}
     \Lambda^* & = - \frac{p}{2} F_0, \\
    F_0 &= \int_{-\pi}^{\pi} \frac{dk}{2 \pi} \frac{1}{2 c \cos k -r} = \frac{f_0}{2c}, \\
    f_0 &= \int_{-\pi}^{\pi} \frac{dk}{2 \pi} \frac{1}{\cos k -D} = - \frac{1}{D \sqrt{1- \frac{1}{D^2}}}, \\
    D &= \frac{r}{2c} = \frac{1}{2 \bar{t}^Z} \left( d_{1,Z} -t_Z^2 d_{2,Z-1} \right) , \\
    \frac{p}{2c}& = D + \frac{ t_Z^2}{ \bar{t}^Z } d_{2,Z-1} ,
\end{align}
where $\bar{t}^Z= \prod_{j=1}^Z t_j$, and
$d_{j,j'}$ are the determinants of tridiagonal matrices which start from $\bar{v}_j$ and end up with $\bar{v}_{j'}$, such that $j \leq j'$. The function $D(\omega)$ determines the dispersion of the bulk problem by virtue of the equation $D (\omega=\epsilon_{k\alpha}) =\cos k$. The edge state equation in the boundary problem reads $s \equiv d_{1,Z-1} =0$. 

Then we express
\begin{align}
& \partial_{\omega}\ln \det \left([G_{0,0}^{(0)}]^{-1} G_{0,1}^{(0)} t_Z \right) = \partial_{\omega}\ln \left( p + \frac{1}{ F_0}\right)  \nonumber \\
=&  \partial_{\omega}\left[\ln \left(  \frac{ t_Z^2}{ \bar{t}^Z } d_{2,Z-1} +D+ \frac{1}{f_0}\right) - \ln s\right].
\end{align}
The argument in the first term is complex valued only above/below branch cuts, which correspond to bands. For each band $\epsilon_{k\alpha}$, lying in the energy range $\epsilon_{ \alpha}^{(-)} <\epsilon_{k \alpha} < \epsilon_{\alpha}^{(+)}$, we can alternatively express this term as an integral over the Bloch momentum $k$ in the Brillouin zone $-\pi < k <\pi$. We notice that
\begin{align}
    \frac{1}{f_0 (\omega+i \eta)} =& - D (\omega) \sqrt{1- \frac{1}{D^2 (\omega+i \eta)}} \\
    =& - i \, \text{sign} [D (\omega) D'(\omega)] D (\omega) \sqrt{\frac{1}{D^2 (\omega)} -1} \nonumber \\
    =& \, i \, \sin k \, \text{sign} \left( \frac{d \epsilon_{k\alpha}}{d k}\right).
\end{align}
For bands with odd $\alpha$, the inequality $\frac{d \epsilon_{k\alpha}}{d k} >0$ holds for $0<k<\pi$, while for even $\alpha$ it holds for $-\pi<k <0$. Therefore, the integration along the upper branch (see Fig.~\ref{fig:winding}) gives
\begin{align}
    & -\frac{1}{\pi} \int_{\epsilon_{\alpha}^{(-)}}^{\epsilon_{\alpha}^{(+)}} d \omega \frac{\partial }{\partial \omega} \ln \left(  \frac{ t_Z^2}{ \bar{t}^Z } d_{2,Z-1} (\omega) +D (\omega)+ \frac{1}{f_0 (\omega + i \eta)}\right) \nonumber \\
    &= \begin{cases} -\frac{1}{\pi} \int_{0}^{\pi} d k \frac{\partial }{\partial k} \ln \left(  \frac{ t_Z^2}{ \bar{t}^Z } d_{2,Z-1} (\epsilon_{k \alpha}) +e^{i k} \right),  & \text{odd} \,\,  \alpha ,\\
    -\frac{1}{\pi} \int_{-\pi}^{0} d k \frac{\partial }{\partial k} \ln \left(  \frac{ t_Z^2}{ \bar{t}^Z } d_{2,Z-1} (\epsilon_{k \alpha}) +e^{i k} \right) ,  & \text{even} \,\,  \alpha .
    \end{cases}
\end{align}
Taking the imaginary part of these expressions and identifying (up to the normalization) $\frac{ t_Z^2}{ \bar{t}^Z } d_{2,Z-1} (\epsilon_{k \alpha}) +e^{i k} \sim \chi_{k \alpha} (1) e^{i k} $ (see Ref.~[\onlinecite{pletyukhov_etal_prb_20}] for details), we obtain the $\alpha$th band's contribution to the invariant in Eq.~\eqref{wn_I}
\begin{align}
    I_{\alpha} &= -\frac{1}{2 \pi i} \int_{-\pi}^{\pi} d k \frac{\partial }{\partial k} \ln \left( \chi_{k \alpha} (1) e^{i k}\right) \\
    &= -1 - \text{wn} [\chi_{k \alpha} (1)],
\end{align}
that is $I_{\alpha}$ is expressed in terms of the winding number of the first Bloch state vector component.

In band gaps, the term 
\begin{align}
    - \frac{1}{\pi} \int d \omega \frac{\partial}{\partial \omega} \ln \left(  \frac{ t_Z^2}{ \bar{t}^Z } d_{2,Z-1} +D+ \frac{1}{f_0} \right)
\end{align}
picks up the  contribution $+1$ from the so called left edge state pole  of the initial system (i.e. with $\kappa<0$) and from the right edge state pole of the shifted system (i.e. with $\tilde{\kappa} >0$). In turn,
the term 
\begin{align}
    \frac{1}{\pi} \int d \omega \frac{\partial}{\partial \omega} \ln s
\end{align}
picks up the  contribution $+1$ from both the left ($\kappa <0$) and right ($\kappa >0$) edge state poles  of the initial system. 

Thus, the invariant defined in Eq.~\eqref{wn_I} in the single channel case amounts to
\begin{align}
    I = \sum_{\alpha=1}^{\nu} I_{\alpha} + \Delta Q_E,
    \label{inv_single_ch}
\end{align}
where $\Delta Q_E$ denotes the difference between numbers of right edge states of the shifted and initial systems summed up over all gaps below the chemical potential $\mu$. Thereby we recover our earlier result expressed in Eqs. (257), (258) of Ref.~[\onlinecite{pletyukhov_etal_prb_20}].

Remarkably, on the basis of \eqref{idL3} we establish the identity
\begin{align}
   \left(  \frac{ t_Z^2}{ \bar{t}^Z } d_{2,Z-1} +D+ \frac{1}{f_0}\right) \left(  \frac{ t_Z^2}{ \bar{t}^Z } d_{2,Z-1} +D-\frac{1}{f_0}\right) =
   \frac{t_Z^2}{\bar{t}^{2Z}} s \tilde{s},
\end{align}
where $\tilde{s}= d_{2,Z}$. It allows us to relate
\begin{align}
& \ln \left(  \frac{ t_Z^2}{ \bar{t}^Z } d_{2,Z-1} +D+ \frac{1}{f_0}\right) - \ln s \nonumber \\
= & \ln \tilde{s} - \ln \left(  \frac{ t_Z^2}{ \bar{t}^Z } d_{2,Z-1} +D- \frac{1}{f_0}\right).
\end{align}
The equation $\tilde{s}=0$ is known to determine both right ($\tilde{\kappa}>0$) and left ($\tilde{\kappa}<0$) edge  states in the shifted system. In turn, $ \frac{ t_Z^2}{ \bar{t}^Z } d_{2,Z-1} +D- \frac{1}{f_0}=0$ holds at the left edge state of the shifted system ($\tilde{\kappa}<0$) and at the right edge of the initial system ($\kappa>0$). Therefore, this observation additionally confirms the result in Eq.~\eqref{inv_single_ch}.

\subsection{Interface charge invariant}
\label{app:single_ch_interface}

In the single-channel case we find
\begin{align}
\Lambda + \frac12 &= - \frac12 \left( \frac{t_Z^2}{\bar{t}^Z} d_{2,Z-1} + D - \frac{1}{f_0} \right) f_0.
\end{align}
Therefore, the contribution to Eq.~\eqref{QI_lambda} from band $\alpha$ may be written as
\begin{align}
\tilde{Q}_{I,\alpha} =&  - \frac{1}{\pi} \int_{\epsilon_{\alpha}^{(-)}}^{\epsilon_{\alpha}^{(+)}} d \omega \,\, \text{Im}  \, \frac{\partial}{\partial \omega}  \ln f_0  \label{bound_term} \\
&  - \frac{1}{\pi} \int_{\epsilon_{\alpha}^{(-)}}^{\epsilon_{\alpha}^{(+)}} d \omega \,\,  \text{Im}  \, \frac{\partial}{\partial \omega}  \ln \left[- \left( \frac{t_Z^2}{\bar{t}^Z} d_{2,Z-1} + D - \frac{1}{f_0} \right) \right. \nonumber \\
& \left. \qquad +\lambda^2  \left( \frac{t_Z^2}{\bar{t}^Z} d_{2,Z-1} + D + \frac{1}{f_0} \right) \right].
\label{bz_term}
\end{align}
The term in Eq.~\eqref{bound_term} receives contributions $-\frac12$ from each band edge. In the term Eq.~\eqref{bz_term} we make the same change of integration variable $\omega\to k$ as discussed in Sec.~\ref{app:single_ch_boundary}. This results in the expression
\begin{align}
& Q_{I,\alpha} +1 \nonumber \\
=& - \frac{1}{2 \pi i} \int_{-\pi}^{\pi} d k \, \frac{\partial}{\partial k} \ln \left[ - e^{-i \varphi_{k \alpha} (1)-i k} + \lambda^2 e^{i \varphi_{k \alpha} (1)+i k}\right] \nonumber \\
=& 1 + \text{wn} [\chi_{k \alpha} (1)] \nonumber \\
& - \frac{1}{2 \pi i} \int_{-\pi}^{\pi} d k \, \frac{\partial}{\partial k} \ln \left[ - 1+ \lambda^2 e^{2 i \varphi_{k \alpha} (1)+2 i k}\right],
\end{align}
where $e^{i \varphi_{k \alpha} (1)}$ is the phase of $\chi_{k \alpha} (1)$. For $\lambda^2 < 1$ the argument in the last line does not encircle the origin, therefore the corresponding contribution vanishes. We finally get
\begin{align}
    Q_{I,\alpha} = \text{wn} \, [\chi_{k \alpha} (1)],
\end{align}
thereby reproducing our earlier result in Eq.~(C35) of Ref.~[\onlinecite{Pletyukhov_etal3_2020}] (assuming the gauge fixed by the condition $\text{Im} \, [\chi_{k \alpha} (Z)]=0$).

\end{appendix}

\end{document}


\title{Supplemental Material: Universal properties of boundary and interface charges in multichannel one-dimensional models without symmetry constraints}

\author{Niclas M\"uller}
\affiliation{Institut f\"ur Theorie der Statistischen Physik, RWTH Aachen, 
52056 Aachen, Germany and JARA - Fundamentals of Future Information Technology}

\author{Kiryl Piasotski}
\affiliation{Institut f\"ur Theorie der Statistischen Physik, RWTH Aachen, 
52056 Aachen, Germany and JARA - Fundamentals of Future Information Technology}

\author{Dante M. Kennes}
\affiliation{Institut f\"ur Theorie der Statistischen Physik, RWTH Aachen, 
52056 Aachen, Germany and JARA - Fundamentals of Future Information Technology}
\affiliation{Max Planck Institute for the Structure and Dynamics of Matter, Center for Free Electron Laser Science, 22761 Hamburg, Germany}

\author{Herbert Schoeller}
\affiliation{Institut f\"ur Theorie der Statistischen Physik, RWTH Aachen, 
52056 Aachen, Germany and JARA - Fundamentals of Future Information Technology}

\author{Mikhail Pletyukhov}
\email[Email: ]{pletmikh@physik.rwth-aachen.de}
\affiliation{Institut f\"ur Theorie der Statistischen Physik, RWTH Aachen, 
52056 Aachen, Germany and JARA - Fundamentals of Future Information Technology}





\maketitle

\label{app:hamiltonians_numerics}
In the following, we list the Hamiltonians used in the numerical examples of section V in the main text.

The $N_c=Z=3$ Hamiltonian used in the calculation of the contour (Fig.~2) is given by
\begin{widetext} 
\begin{align}
    v_1 &= 
    \begin{pmatrix}
    -0.4944582 &          0.57434879+0.76768438 i & 0.38010216-0.3227788i \\
    0.57434879-0.76768438i &  0.17028087 &         -0.12163262+0.08683064 i \\
    0.38010216+0.3227788 i & -0.12163262-0.08683064 i & -0.1872981 
    \end{pmatrix}, \\
    v_2 &= 
    \begin{pmatrix}
    0.54582815 &         -0.36852351-0.04013338 i &  0.17708711-0.15488239 i \\
    -0.36852351+0.04013338 i &  0.98698514 &          0.23707274+0.03260454 i \\
    0.17708711+0.15488239 i &  0.23707274-0.03260454 i & -0.89519627
    \end{pmatrix}, \\
    v_3 &= 
    \begin{pmatrix}
    0.09034543 &        -0.0529429 +0.66493465 i & -0.40929606+0.50975004 i \\
    -0.0529429 -0.66493465 i & 0.62717073 &         -0.70806207-0.46843796 i \\
    -0.40929606-0.50975004 i & -0.70806207+0.46843796 i & -0.84868633       
    \end{pmatrix}, \\
    t_1 &= 
    \begin{pmatrix}
    0.24106425 + 0.75954503 i & 0.39951955-0.83987445 i &  -0.70939688 + 0.59471041 i \\
    0.08062616 - 0.62072405 i &  -0.64056866 + 0.46943078 i & 0.69665083 + 0.56323945 i \\
    -0.15140389 + 0.62351802 i & -0.4158642  + 0.61600451 i &  -0.47909182 - 0.468285 i
    \end{pmatrix}, \\
    t_2 &= 
    \begin{pmatrix}
    0.95472693 + 0.27556248 i & 0.79839864 + 0.0917834 i &  0.71196756 + 0.51063241 i \\
    0.3859219  - 0.06135333 i & 0.49234291 - 0.03195245 i &  -0.8642979  - 0.31707168 i \\
    0.35721307 - 0.61547973 i & 0.68259592 + 0.97858545 i &  -0.45682217 - 0.87495712 i
    \end{pmatrix}, \\
    t_3 &= 
    \begin{pmatrix}
    0.53283463 - 0.01128521 i &  -0.15628193 + 0.98906666 i & 0.03970061 - 0.72966195 i \\
    0.36798589 + 0.38419227 i & 0.42645765 + 0.71823889 i & 0.70707806 - 0.69956256 i \\
    0.84265784 - 0.19864427 i &  -0.25684613 + 0.07105722 i &  -0.49867149 - 0.33685095 i
    \end{pmatrix}. 
\end{align}
\end{widetext}

The $N_c=3, Z=4$ Hamiltonian used in the calculation of the shifted boundary charge invariant example (Fig.~3) is given by
\begin{widetext}
\begin{align}
    v_0 &= 
    \begin{pmatrix}
     0.90918084 &  -0.24572285-0.66887345i & 0.07488023+0.43240801i \\
     -0.24572285+0.66887345i & 0.56819934 & -0.39061585-0.22247052i \\
      0.07488023-0.43240801i & -0.39061585+0.22247052i & -0.17051496
    \end{pmatrix}, \\
    \delta_v &= 
    \begin{pmatrix}
     0.02774156 &        0.2859403 -0.01955917i & -0.28670403-0.09992215i \\
  0.2859403 +0.01955917i &-0.07877183 &         -0.64531684+0.08964048i \\
 -0.28670403+0.09992215i &-0.64531684-0.08964048i &-0.93476734
    \end{pmatrix}, \\
    t_0 &=
    \begin{pmatrix}
    -0.88944725-0.76770169i &-0.32851701+0.4908495i &   0.17681831+0.44297607i \\
  0.27956042-0.15460026i & 0.69664396-0.80789705i& -0.06244329-0.36763672i \\
  0.33002784+0.57796944i &-0.9874527 +0.92545997i & 0.27819225-0.47315019i
    \end{pmatrix}, \\
    \delta_t &= 
    \begin{pmatrix}
    -0.58603661-0.1884779i &  0.37426473-0.12101691i & 0.34827591+0.0989828i \\
 -0.66360162+0.84043632i & 0.29328977+0.06031052i &-0.23946274+0.35388921i \\
  0.49854692+0.282901i &   0.00510658-0.49200114i & 0.9002086 -0.00963393i
    \end{pmatrix}.
\end{align}
\end{widetext}

The $N_c=4, Z=3$ Hamiltonian used in the calculation of the potential impurity interface charge example (Fig.~4) is given by
\small
\begin{widetext}

\begin{align}
    v_1 &=
    \begin{pmatrix}
    -0.60687141 &       0.12885404-0.28928961i &  0.17126335+0.82789126i &
  -0.01562436-0.42902267i \\
  0.12885404+0.28928961i & -0.23745729 &         -0.02768967-0.16012022i &
  -0.5607436 +0.09055274i \\
  0.17126335-0.82789126i & -0.02768967+0.16012022i & 0.99461468 &
   0.22963881-0.35765889i \\
 -0.01562436+0.42902267i & -0.5607436 -0.09055274i &  0.22963881+0.35765889i &
  -0.74220396        
    \end{pmatrix}, \\
    v_2 &=
    \begin{pmatrix}
    -0.71744785 &         -0.02397275-0.16690444i & 0.21099762-0.01700339i &
  -0.13142301+0.38309352i \\
 -0.02397275+0.16690444i &  0.76671614 &         0.23552629-0.32028269i &
   0.85170206+0.07966246i \\
  0.21099762+0.01700339i & 0.23552629+0.32028269i & 0.44805615 &
   0.76446223+0.41996331i \\
 -0.13142301-0.38309352i & 0.85170206-0.07966246i & 0.76446223-0.41996331i &
  -0.28404753        
    \end{pmatrix}, \\
    v_3 &= 
    \begin{pmatrix}
    0.41620817 &        -0.36871262+0.96755171i & 0.35036057+0.25990987i
&  -0.2083217 +0.41540971i \\
 -0.36871262-0.96755171i & 0.05247915 &         -0.32631566+0.48538255i &
  -0.21221584-0.31134165i \\
  0.35036057-0.25990987i  & -0.32631566-0.48538255i & 0.42443833 &
   0.49352163+0.31176955i \\
 -0.2083217 -0.41540971i & -0.21221584+0.31134165i & 0.49352163-0.31176955i &
  -0.53993851      
    \end{pmatrix}, \\
    t_1 &=
    \begin{pmatrix}
    -0.40522514+0.28553213i & -0.65262785+0.30669345i & 0.16121097-0.51314027i
&   0.83699647+0.99655913i \\
 -0.67139569+0.21442743i & 0.17163123+0.12591744i & -0.63191814-0.13116329i &
   0.92153573-0.25782209i \\
 -0.61992322+0.28293355i & -0.35989309+0.18055828i & -0.36809224+0.64191059i &
  -0.39694743+0.23910722i \\
 -0.5400065 -0.79888228i & 0.30690518-0.03598716i & 0.7106456 -0.08485762i &
  -0.84147776-0.27194106i
    \end{pmatrix}, \\
    t_2 &=
    \begin{pmatrix}
    0.76176153-0.77515945i &  0.80689694+0.53367223i & 0.77512597+0.81955277i
&   0.39171962+0.3681039i  \\
 -0.86282357+0.87501381i &  0.84833074+0.75070133i & -0.31790326+0.44673359i &
  -0.6401189 +0.09685302i \\
  0.45167651-0.0597095i &   0.85496192+0.63632698i & -0.79501046+0.57622953i &
  -0.15800389-0.06831002i \\
 -0.76541453-0.31212747i & -0.23971583-0.11841097i & -0.7392837 +0.60364822i &
  -0.98611912-0.02725172i
    \end{pmatrix}, \\
    t_3 &= 
    \begin{pmatrix}
    -0.51495422-0.68189032i & -0.07052153-0.75208249i & -0.29828693+0.3386923i &
  -0.22381824-0.19244218i \\
  0.72135027-0.90520099i & -0.12802552-0.18513815i &  0.78158383-0.76246528i &
  -0.51494   +0.65168529i \\
  0.14336053+0.64694445i &  0.73489241+0.05943474i & 0.35838066+0.42942297i &
  -0.15150548-0.4830256i \\
 -0.11175007-0.20852716i &  0.45727396-0.80496684i & 0.91112461-0.73402748i &
   0.47394746+0.12491548i
    \end{pmatrix}. 
\end{align}
\end{widetext}
\normalsize

The $N_c=5, Z=5$ Hamiltonian used in the calculation of the link-weakened interface charge example (Fig.~5) is given by

\tiny
\begin{widetext}
\begin{align}
    v_1 &= 
    \begin{pmatrix}
        -0.84739301   &       0.15192162-0.35657652i & 0.39340146-0.72879555i &
   0.05229739-0.32752712i & -0.07163662-0.39608318i \\
  0.15192162+0.35657652i &  0.17513174  &       -0.59635007+0.77099924i &
  -0.55614948+0.54347114i & -0.25198908-0.09196395i \\
  0.39340146+0.72879555i & -0.59635007-0.77099924i & -0.30659615 &
   0.38055026+0.24921779i & 0.57581586+0.12289488i \\
  0.05229739+0.32752712i & -0.55614948-0.54347114i & 0.38055026-0.24921779i &
  -0.48690864  &        0.03716538-0.1713455i  \\
 -0.07163662+0.39608318i & -0.25198908+0.09196395i & 0.57581586-0.12289488i &
   0.03716538+0.1713455i &  0.40536641       
    \end{pmatrix}, \\
    v_2 &= 
    \begin{pmatrix}
        -0.95387331   &       0.01899544-0.52384378i & -0.06753597+0.23454486i &
   0.08562934+0.21877701i &  0.39594011+0.18401971i \\
  0.01899544+0.52384378i &  0.33904945  &       -0.12323575+0.58328219i &
  -0.26313186-0.16264879i & 0.24225949+0.57721432i \\
 -0.06753597-0.23454486i & -0.12323575-0.58328219i & -0.47888902 &
   0.02524494+0.29215537i &  0.38316456-0.06532999i \\
  0.08562934-0.21877701i & -0.26313186+0.16264879i &  0.02524494-0.29215537i &
   0.18735596 &         0.25849069-0.08107532i \\
  0.39594011-0.18401971i & 0.24225949-0.57721432i & 0.38316456+0.06532999i &
   0.25849069+0.08107532i & -0.42200492        
    \end{pmatrix}, \\
    v_3 &= 
    \begin{pmatrix}
        -0.01520067  &       -0.16572919+0.2040341i &  0.19093657+0.25573464i
&  -0.09632023+0.4730379i  & 0.1955689 +0.06831824i \\
 -0.16572919-0.2040341i &  0.95158912  &        0.17033616-0.64845533i &
  -0.19121934+0.82801133i & -0.5146001 +0.13980757i \\
  0.19093657-0.25573464i &  0.17033616+0.64845533i & -0.79367107 &
   0.30450466-0.79284896i &  0.25800025+0.18407783i \\
 -0.09632023-0.4730379i &  -0.19121934-0.82801133i &  0.30450466+0.79284896i &
  -0.5404087        &   0.2743869 +0.22916055i \\
  0.1955689 -0.06831824i & -0.5146001 -0.13980757i & 0.25800025-0.18407783i &
   0.2743869 -0.22916055i &  0.83056931       
    \end{pmatrix}, \\
    v_4 &= 
    \begin{pmatrix}
        -0.56018624 &        -0.30849839+0.13792053i & -0.09705989-0.20756389i &
  -0.01036744-0.45287618i & -0.42858509-0.73719311i \\
 -0.30849839-0.13792053i & -0.70401226  &        0.74535439-0.05772667i &
  -0.00306166+0.4168619i &  -0.88935293-0.14014358i \\
 -0.09705989+0.20756389i &  0.74535439+0.05772667i &  0.41833938
&  -0.91932691-0.18471939i & 0.72000204+0.0246125i \\
 -0.01036744+0.45287618i & -0.00306166-0.4168619i & -0.91932691+0.18471939i &
   0.49192261  &       -0.62016451+0.07836037i \\
 -0.42858509+0.73719311i & -0.88935293+0.14014358i &  0.72000204-0.0246125i &
  -0.62016451-0.07836037i &  0.96386105        
    \end{pmatrix}, \\
    v_5 &= 
    \begin{pmatrix}
-0.57167868  &       -0.13678641+0.562379i &  -0.76211957+0.13162859i &
  -0.06327828-0.02382753i &  0.04930314-0.01028572i \\
 -0.13678641-0.562379i &   0.04751132  &       -0.34385889-0.69313275i &
   0.655804  +0.85336297i & 0.75116565-0.00889398i \\
 -0.76211957-0.13162859i & -0.34385889+0.69313275i & -0.254086 & 
  -0.04476031-0.13983655i & -0.1657417 -0.11132861i \\
 -0.06327828+0.02382753i & 0.655804  -0.85336297i & -0.04476031+0.13983655i &
   0.38239353   &       0.13682864-0.11788784i \\
  0.04930314+0.01028572i & 0.75116565+0.00889398i & -0.1657417 +0.11132861i &
   0.13682864+0.11788784i & 0.04561698        
    \end{pmatrix}, \\
    t_1 &= 
    \begin{pmatrix}
        0.80047348-0.85591018i & 0.23149352+0.44369256i & -0.23002532+0.17816297i &
  -0.39921379+0.84223912i & -0.014936  -0.89550549i  \\
  0.90443249-0.4340823i  &  0.39078169-0.26052319i  & 0.32689639+0.48807573i &
   0.3168691 +0.1521063i  & -0.90555087+0.87287847i \\
  0.89201688-0.27274039i & -0.37060051+0.76101001i  &-0.37897776-0.07010669i &
  -0.16707652+0.63068628i & -0.22471651-0.31333483i \\
 -0.03780648-0.53863946i &  0.21728634-0.18827376i  &-0.55491977-0.93217545i &
   0.19125778-0.74817493i & -0.02959721+0.07231875i \\
 -0.53426739+0.6217711i &   0.63639065-0.14903255i  & 0.85241615+0.4489814i &
  -0.14613378+0.748662i  &   0.2936694 +0.58832297i
    \end{pmatrix}, \\
    t_2 &= 
    \begin{pmatrix}
        0.46992338+0.16509969i & -0.30436618+0.47388353i  & 0.49098679+0.98164082i &
  -0.54695302+0.51214401i & -0.37586555+0.17905454i \\
  0.02707689+0.10524192i &  0.93233687+0.7648932i &  -0.58810739+0.40664544i &
   0.5716518 -0.83174955i & -0.65173778+0.206885i  \\
  0.55975019-0.53669275i &  0.26586444+0.29762377i & -0.49429293-0.1435033i &
   0.92650894-0.84611881i & -0.70455786-0.48409096i \\
  0.59037792-0.54091199i &  0.85788389+0.24560461i &  0.47476284+0.49301538i &
  -0.69627716-0.68994854i &  0.55997991+0.56031605i \\
  0.63311455-0.39776219i &  0.8913214 +0.8528227i &   0.17073605-0.13321926i &
  -0.44129796+0.97878991i &  0.91049575-0.51784284i
    \end{pmatrix}, \\
    t_3 &= 
    \begin{pmatrix}
        -0.56639054+0.95027621i & -0.52393862+0.04569261i & -0.09484941-0.76356329i &
  -0.5153158 +0.49819619i &  0.61664959-0.59327422i \\
 -0.30669642+0.68436901i &  0.56985208-0.00448519i &  0.67788516-0.30013223i &
  -0.53975312-0.525708i &    0.88646412-0.47259366i \\
 -0.08189609+0.6883885i &  -0.67302818-0.04206554i &  0.65960854-0.27529002i &
  -0.8154709 +0.37974371i &  0.66780298+0.04558129i \\
  0.37393171-0.70087709i & -0.60133584+0.6991957i &   0.3181342 -0.05591992i &
   0.56664716+0.8622991i &   0.82282491-0.5321042i  \\
  0.30922302+0.37900218i & -0.49970823+0.38016214i &  0.1028031 +0.44109531i &
   0.41420559-0.97487061i &  0.52013369+0.86341996i
    \end{pmatrix}, \\
    t_4 &= 
    \begin{pmatrix}
        -0.89886607+0.06847595i & -0.9166851 +0.49653758i & -0.53174656+0.89220242i &
  -0.72362803+0.11476783i &  0.74469864+0.499776i  \\
 -0.48299081-0.49399424i & -0.27648116+0.01925675i & -0.16821297-0.89373263i &
   0.33594719-0.16372056i &  0.2663226 +0.77684248i \\
  0.28205414+0.69985943i &  0.59756142-0.11459214i & -0.51297052+0.90461684i &
  -0.10778088-0.94646817i &  0.33354676+0.32009903i \\
  0.29740809-0.66640539i &  0.18472455+0.95000883i & -0.21045865+0.17550876i &
   0.9077019 -0.77096572i & -0.90006669+0.19303063i \\
 -0.14091869+0.60143292i &  0.2484769 +0.50801011i &  0.04669818+0.85133981i &
   0.49710369-0.8538442i &   0.89424029+0.01437209i
    \end{pmatrix}, \\
    t_5 &= 
    \begin{pmatrix}
        -0.92432482+0.03243803i & -0.3920592 +0.52241285i &  0.96245486+0.14039386i &
   0.30996196+0.84366345i &  0.28303399-0.22646471i \\
  0.60217582+0.84160544i & -0.72616155-0.05684327i &  0.01138053-0.50204865i &
   0.08245395-0.85951913i &  0.13246862-0.09641901i \\
  0.73965747+0.30292888i &  0.75622971+0.28772264i &  0.18217603-0.57157653i &
  -0.74675744-0.34726472i &  0.39416056+0.7220469i  \\
  0.95941996-0.67000463i & -0.7789017 +0.20999963i &  0.75104373-0.19257495i &
  -0.97048715-0.63885318i &  0.9377695 +0.83096135i \\
  0.28838462+0.43587651i & -0.06389155-0.16687494i & -0.54631375+0.29305552i &
  -0.47303696+0.519649i &   -0.59851634+0.94455056i
    \end{pmatrix}. 
\end{align}
\end{widetext}
\normalsize